\documentclass[
 aps,
 11pt,
 final,
 notitlepage,
 oneside,
 twocolumn,
 nobibnotes,
 nofootinbib,
 superscriptaddress,
 noshowpacs,
 centertags,
 showkeys]
{revtex4}

\def\squareforqed{\hbox{\rlap{$\sqcap$}$\sqcup$}}

\def\sq{\ifmmode\squareforqed\else{\unskip\nobreak\hfil
\penalty50\hskip1em\null\nobreak\hfil\squareforqed
\parfillskip=0pt\finalhyphendemerits=0\endgraf}\fi}

\def\utw{\smash{\rlap{\lower5pt\hbox{$\sim$}}}}

\def\udtw{\smash{\rlap{\lower6pt\hbox{$\approx$}}}}

\def\diameter{{\ifmmode\mathchoice
{\ooalign{\hfil\hbox{$\displaystyle/$}\hfil\crcr
{\hbox{$\displaystyle\mathchar"20D$}}}}
{\ooalign{\hfil\hbox{$\textstyle/$}\hfil\crcr
{\hbox{$\textstyle\mathchar"20D$}}}}
{\ooalign{\hfil\hbox{$\scriptstyle/$}\hfil\crcr
{\hbox{$\scriptstyle\mathchar"20D$}}}}
{\ooalign{\hfil\hbox{$\scriptscriptstyle/$}\hfil\crcr
{\hbox{$\scriptscriptstyle\mathchar"20D$}}}}
\else{\ooalign{\hfil/\hfil\crcr\mathhexbox20D}}%
\fi}}

\begin{document}
\selectlanguage{English}

% ¬несите сюда свои ключевые слова

%\keywords{stars: magnetic field---stars: chemically peculiar---stars: individual: $\alpha^2$\,CVn}

% ƒанные строчки заполн€ютс€ техническим редактором, оставьте их закомментированными
%
%\ydk{}
%\titlerunning{}
%\authorrunning{}
%\toctitle{}
%\tocauthor{}
%\received{}  \revised{}

% «аголовок статьи

\noindent Butenko M.A., Khoperskov A.V. Galaxies with ``Rows'': A New Catalog // Astrophysical Bulletin, 2017, vol.72, No.3, 232--250 \\
 https://doi.org/10.1134/S1990341317030130

\

\

\title{GALAXIES WITH УROWSФ: A NEW CATALOG}

% ѕервый автор, e-mail первого автора, место работы первого автора.
% Ќазвание —јќ –јЌ можно ввести с помощью команды \saonamer

\author{\firstname{M.~A.}~\surname{Butenko}}
\email{maria_butenko@volsu.ru}
\affiliation{Volgograd State University, Volgograd, 400062 Russia}

 \author{\firstname{A.~V.}~\surname{Khoperskov}}
 % «десь разбиение на строки осуществл€етс€ автоматически или командой \\
 \email{khoperskov@volsu.ru}
 \affiliation{Volgograd State University, Volgograd, 400062 Russia}%

% ратка€ аннотаци€

\begin{abstract}
Galaxies with УrowsФ in Vorontsov-VelyaminovТs terminology stand out among the variety of
spiral galactic patterns. A characteristic feature of such objects is the sequence of straight-line segments
that forms the spiral arm. In 2001 A. Chernin and co-authors published a catalog of such galaxies which
includes~204 objects from the Palomar Atlas. In this paper, we supplement the catalog with 276 objects
based on an analysis of all the galaxies from the New General Catalogue and Index Catalogue. The total
number of NGC and IC galaxies with rows is~406, including the objects of Chernin et al. (2001). The use
of more recent galaxy images allowed us to detect more УrowsФ on average, compared with the catalog
of Chernin et al. When comparing the principal galaxy properties we found no significant differences
between galaxies with rows and all S-typeNGC/IC galaxies.We discuss twomechanisms for the formation
of polygonal structures based on numerical gas-dynamic and collisionless N-body calculations, which
demonstrate that a spiral pattern with rows is a transient stage in the evolution of galaxies and a system
with a powerful spiral structure can pass through this stage. The hypothesis of A.~Chernin et al. (2001) that
the occurrence frequency of interacting galaxies is twice higher among galaxies with rows is not confirmed
for the combined set of~480 galaxies. The presence of a central stellar bar appears to be a favorable factor
for the formation of a system of УrowsФ.
\end{abstract}

%  оманда \maketitle оформл€ет заголовок статьи

\maketitle

%Ќазвани€ разделов статьи ввод€тс€ с помощью команды \section
%ѕодразделы - c помощью команд \subsection и \subsubsection

\section{INTRODUCTION}\label{section:introduction}

% ƒалее вводитс€ сам текст статьи. –екомендуетс€ обратить внимание на оформление
% математических переменных, ссылок, тире и т.д.
% ќбразцы вставки в текст таблиц, рисунков, формул приведены далее в тексте.

Straight-line segments that form the spiral pattern
can be seen in the images of some spiral
galaxies. Vorontsov-Velyaminov~\cite{Vorontsov-Velyaminov_1964}, must be the first
to point out this feature, calling the said segments
УrowsФ. Such rather long and practically straightline
segments form uneven, but almost regular spiral
arms, and these structures are often called polygonal
arms~\cite{Chernin_geksagon_2001}. The M101 and M51 galaxies are typical
examples. A. D.~Chernin with coauthors compiled
a catalog of galaxies with УrowsФ including 204
objects~\cite{Chernin_etal_2001, Chernin_etal_2000}. An analysis of such structures revealed
that they have the following properties~\cite{Chernin_etal_2001}:
\begin{enumerate}
	\item The length $L$ of the straight-line segment (УrowФ)
linearly increases with galactocentric distance $d$, $d = (1 \pm  0.11) L$.
	\item The angle between the two adjacent segments
is close to $\alpha = 120^\circ$ (with a standard deviation of $10^\circ$).
	\item Straight-line segments can be subdivided into two types: those that form rather regular global structure of the spiral arm and oneЦtwo segments that
do not form the global spiral pattern.
	\item УRowsФ are observed mostly in galaxies of late morphological types Sbc--Scd.
	\item Straight-line segments occur more often in interacting galaxies.
	\item The average number of УrowsФ in a galaxy is \mbox{$N = 3$}.
	\item Galaxies with УrowsФ are rather rare objects making
up for about $\sim 7$\%  of all spiral galaxies with welldefined
spiral arms.
\end{enumerate}
Note that these results are based on an analysis of
photographic plates and Palomar Atlas images.
In this paper we report the results of an analysis
of our catalog of galaxies with polygonal structures,
which lists 276 objects not included into the earlier
published catalog~\cite{Chernin_etal_2001}. When combined, these catalogs
include all NGC and IC galaxies with УrowsФ.

\section{SAMPLE OF GALAXIES WITH УROWSФ}
\subsection{General Characterization}\label{subsection_Character}

\begin{figure*}[!htb]
 \setcaptionmargin{5mm} \onelinecaptionstrue \captionstyle{normal}
 \includegraphics[scale=0.8]{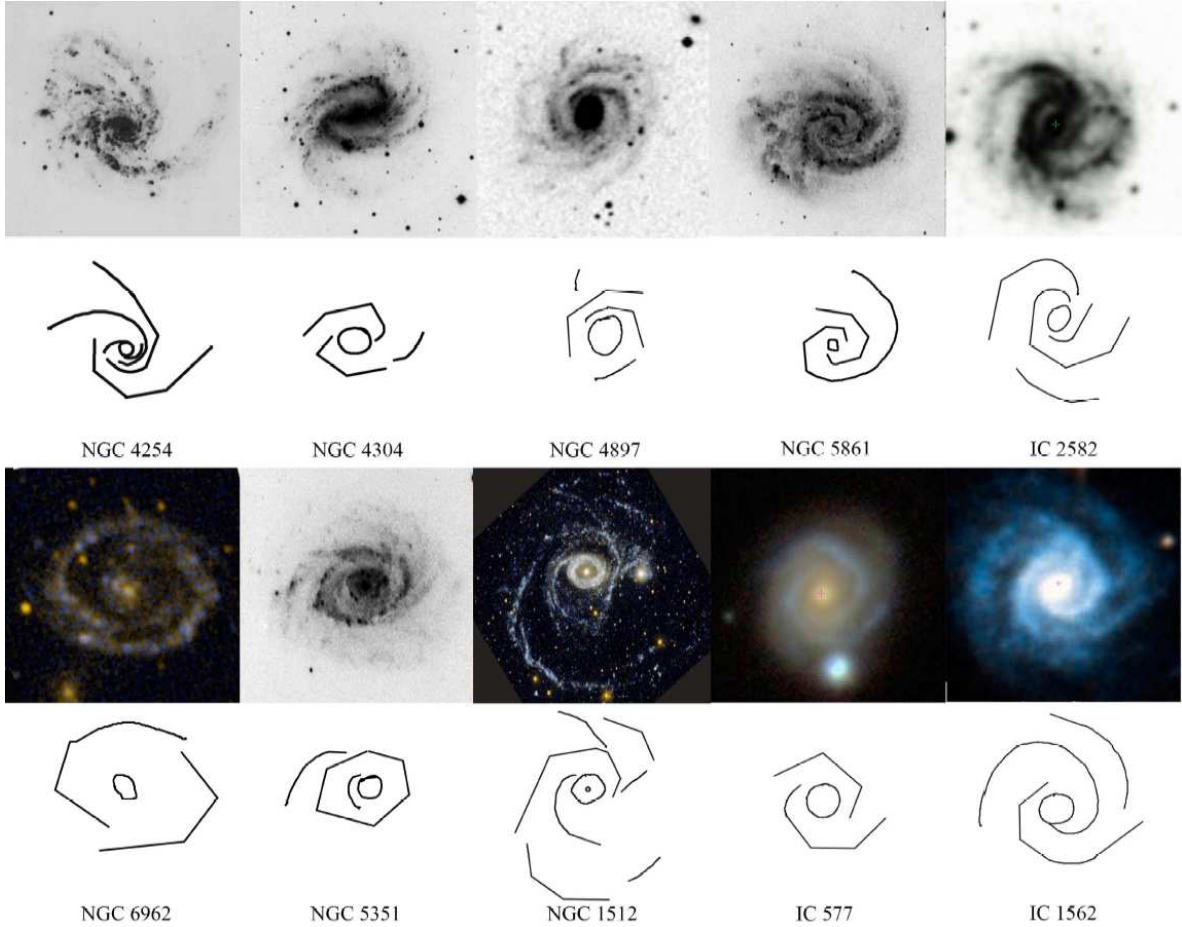} %{Butenko_fig01.eps}    % —помощью параметра scale можно мен€ть размер рисунка
 \caption{Examples of galaxies with УrowsФ forming various geometric structures.}
 \label{fig1:Example_PoligonStr}
\end{figure*}

When searching for galaxies with УrowsФ (straightline
arm segments) we inspected more than 30 000 images of spiral galaxies taken in different parts
of the spectrum and adopted from astronomical
databases. This includes all spiral galaxies in the
NGC/IC catalogs. Our analysis revealed 276 galaxies
with straight-line segments in addition to the
204 galaxies with УrowsФ listed in the catalog of
Chernin et al.~\cite{Chernin_etal_2001}. We imposed the following additional
constraints: declination $\delta>-45^\circ$, objects should be
brighter than $15^m$, inclination of the galaxy to the
line of sight $i<70^\circ$, distance less than 200~Mpc, and $R_{25}\lesssim 30$~kpc.
When compiling our sample we
used images from all available sources including
DSS, SDSS, GALEX, 2MASS, and HST.We show
some examples of galaxies with УrowsФ in~\ref{fig1:Example_PoligonStr}. We
also identify so-called hexagonal structures~\cite{Chernin_geksagon_2001, Buta_Combes_1996}, which form almost ring-like features in the galactic
disk (NGC~4736, NGC~5351, NGC~6962, NGC~7329, IC~1764, IC~4688).

% ¬ставл€ем рисунок, занимающий обе колонки текста (всю ширину листа).
% –исунки и таблицы рекомендуетс€ вставл€ть сразу после абзаца, в котором
% они упоминаютс€ в первый раз. ѕри вставке рисунок можно масштабировать с помощью
% параметров scale, width. –азмер надписей на рисунке в итоге должен быть примерно таким,
% как в тексте. Ќадписи одного типа (например, размер цифр по ос€м x и y)
% должны быть идентичными на одном рисунке и примерно одинаковыми дл€ разных рисунков.
%
% »дентичные рисунки должны выгл€деть одинаково: одинаковые пропорции (ширина-высота),
% шрифты, размеры надписей.

 %Fig 1

In some objects polygons can be detected in various
spectral ranges, however, there are exceptions.
Polygonal structures in 2MASS images could be
found only in the NGC~4254, NGC~5156, NGC~5351, NGC~5653, NGC~5968, NGC~6035, NGC~6691, NGC~7678, IC~1142, IC~2627, IC~4219, IC~4359, IC~4444, IC~4567, IC~4646, IC~4836, IC~4839, IC~5325 galaxies.

We performed primary selection by examining
galaxy images from astronomical databases.We then
processed the images of such galaxies as described
in~\cite{Butenko_2015}. We transformed the image of a galaxy with rows
to the Уface-onФ form ($i\approx 0$) by rotating it for the
major axis to coincide with one of the coordinate axis
and stretching the image along the direction of the
minor axis assuming that the disk is infinitely thin.
We then superimposed onto the transformed galaxy
image lines tracing the geometry of УrowsФ. We then
determined for each galaxy: the length $L_i$ of $ith$ УrowФ,
distance $d_i$ from the galactic center to the edge of the
УrowФ, and angle $\alpha_i$ between the adjacent straightline
segments.
To estimate the error of this method, we performed
series of simulations by taking pictures of schematic
images of various polygonal structures at different
angles $i$ and transforming such images to the УfaceonФ
form.
At fixed $i$ the errors of $L$ and $\alpha$ estimates depend
strongly on the geometry of the polygonal structure
and the location of the particular УrowФ in the plane
perpendicular to the line of sight. Depending on
the location of УrowsФ the error of estimated $L$ and $\alpha$ lies within $0 \leqslant \varepsilon \leqslant \varepsilon_{\max}$. Our simulations yield $\varepsilon_{\max}^{(L)} = 12$\% and $\varepsilon_{\max}^{(\alpha)} = 6$\% for $i=60^\circ$, decreasing down to $\varepsilon_{\max}^{(L)} = 9$\% and $\varepsilon_{\max}^{(\alpha)} = 3$\% for $i=30^\circ$. In
the case of random location and orientation of the
УrowФ the error of its parameter estimates is of about $\varepsilon_{\max}/2$.
When averaged over all simulations and
all УrowsФ the error does not exceed 5\% for linear
measurements. A similar averaging for angles yields $\delta \alpha \approx 3$\%.
Hence within the framework of the above
approach the errors of linear and angle parameters do
not exceed the errors of initial observational data even
if УrowsФ are identified by eye.

As a result, we found 276 objects in addition to
those listed in the catalog of Chernin et al.~\cite{Chernin_etal_2001}.~\ref{galaxy_table} gives the following information about our newly
discovered objects: (1) name of the galaxy according
to NGC or IC; (2) morphological type; (3) integrated $B_0$,magnitude corrected for Galactic extinction
and inclination to the line of sight; (4) heliocentric
radial velocity $V_0$, (5) absolute magnitude $M_B$ (computed with $H_0 = 75$ km $s^{-1}$ $Mps^{-1}$); (6) HI
mass-to-luminosity ratio $M_{HI}/L_B$ (in solar units), (7) number of УrowsФ found in the galaxy; (8) reference
to the main image of the galaxy where the
straight-line segments of the spiral structure were
found that was used to count their number, measure
the linear size and angles between the УrowsФ; (9) references
to other images where straight-line УrowsФ
can be found, and (10) information about whether the
galaxy exhibits any signs of interaction. The parameter
values listed in columns (2)Ц(6) of \ref{galaxy_table} вз€ты из LEDA are
adopted from HyperLeda (See \cite{Makarov_Prugniel_Terekhova_etal_2014} and http://leda.univ-lyon1.fr/).

%’арактеристики галактик вз€ты из электронных баз данных LEDA и NED.

% ¬ставл€ем длинную таблицу. »спользуйте это окружение (longtable) только в том случае,
% если таблицу невозможно поместить на одну страницу.

%\section{√алактики с <<вереницами>>}
%
%\onecolumngrid
\setcaptionwidth{\linewidth}
\setcaptionmargin{0mm} %
\onelinecaptionstrue
\captionstyle{normal}
\begin{longtable*}{c|l|c|c|c|c|c|c|c|c}
	\caption{Galaxies with УrowsФ \label{galaxy_table}} \\
	\hline
	NGC/IC  & Type    &  $B_0$  & $V_0$, kms$^{-1}$  & $M_B$  & $M_{HI}/L_B$  & Number   & Main & Other 	& Inter- \\
	&		 &	       &		      &	       &		       & of УrowФ & Reference$^*$ & Reference$^*$ & action$^{**}$  \\
	\hline		
	(1)		&	(2)	 &	  (3)  &		(4)   &	  (5)  &		(6)    & (7) 	 &   (8)      & (9)  	   & (10)  		  \\
	\hline
	\endfirsthead
	\caption{(Contd.)} \\
	\hline
	NGC/IC  & Type    &  $B_0$  & $V_0$, kms$^{-1}$  & $M_B$  & $M_{HI}/L_B$  & Number   & Main & Other 	& Inter- \\
	&		 &	       &		      &	       &		       & of УrowФ & Reference$^*$ & Reference$^*$ & action$^{**}$  \\
	\hline
	(1)		&	(2)	 &	  (3)  &		(4)   &	  (5)  &		(6)    & (7) 	 &   (8)      & (9)  	   & (10)  		  \\
	\hline
	\endhead
	\hline
	\endfoot
	\endlastfoot
	0010 	 & 	Sbc  & 	 12.74	 & 	 6803  & 	-22.18	 & 	 0.18 	 & 	5  &  B   &  C, R 	&   			 \\
	0060 	 & 	Sc   & 	 14.66	 & 	 11809  & 	-21.52	 & 	    	 & 	5  &  SD  &  B 		&  asym		 \\
	0099 	 & 	Sc 	 & 	 13.71	 & 	 5310  & 	-20.74	 & 	 0.86 	 & 	3  &  B   &  SD 	&   			 \\
	0157 	 & 	SABb & 	 10.40	 & 	 1654  & 	-21.41	 & 	 0.17 	 & 	4  &  SD  &  B  	&   			 \\
	0165 	 & 	Sbc  & 	 13.50	 & 	 5889  & 	-21.13	 & 	 0.29 	 & 	5  &  G   &  B 		&   			 \\
	0191 	 & 	SABc & 	 13.75	 & 	 6076  & 	-20.96	 & 	    	 & 	4  &  SD  &  B 		&  ARP127 		 \\
	0201 	 & 	Sc   & 	 13.35	 & 	 4386  & 	-20.66	 & 	 0.25 	 & 	4  &  SD  &  B 		&  M 			 \\
	0214 	 & 	SABc & 	 12.49	 & 	 4535  & 	-21.65	 & 	 0.14 	 & 	5  &  HST &  SD, B 	&   			 \\
	0234 	 & 	SABc & 	 12.91	 & 	 4449  & 	-21.16	 & 	 0.21 	 & 	9  &  SD  &  C, B 	&   			 \\
	0255 	 & 	Sbc  & 	 12.21	 & 	 1597  & 	-19.55	 & 	 0.73 	 & 	3  &  B   &  C 		&   			 \\
	0268 	 & 	Sbc  & 	 13.00	 & 	 5479  & 	-21.49	 & 	 0.46 	 & 	2  &  SD  &  B 		&   			 \\
	0289 	 & 	SBbc & 	 11.39	 & 	 1630  & 	-20.25	 & 	 1.16 	 & 	2  &  G   &  B 		&  VV484, M, Grp  \\
	0300 	 & 	Scd  & 	 8.40	 & 	 165  & 	-18.14	 & 	 1.59 	 & 	6  &  G   &  B 		&   			 \\
	0521 	 & 	Sbc  & 	 12.40	 & 	 5023  & 	-21.90	 & 	 0.15 	 & 	6  &  SD  &  B, C 	&   			 \\
	0578 	 & 	Sc   & 	 11.11	 & 	 1628  & 	-20.59	 & 	 0.33 	 & 	3  &  B   &  C 		&   			 \\
	0685 	 & 	Sc   & 	 11.58	 & 	 1360  & 	-19.45	 & 	 0.63 	 & 	1  &  B   &  C, R 	&   			 \\
	0753 	 & 	SABc & 	 12.32	 & 	 4902  & 	-21.99	 & 	 0.30 	 & 	7  &  B   &  C 			&   			 \\
	0783 	 & 	Sc   & 	 12.43	 & 	 5192  & 	-22.01	 & 	 0.14 	 & 	4  &  G   &  SD, B, C 	&   			 \\
	0799 	 & 	SBa  & 	 13.96	 & 	 5837  & 	-20.66	 & 	 0.31 	 & 	2  &  SD  &  B, C 		&   			 \\
	0800 	 & 	Sc 	 & 	 14.03	 & 	 5934  & 	-20.63	 & 	 0.38 	 & 	2  &  SD  &  B 			&   			 \\
	0877 	 & 	SABc & 	 11.83	 & 	 3913  & 	-21.95	 & 	 0.30 	 & 	5  &  B   &  G, C 		&   			 \\
	0887 	 & 	SABc & 	 13.02	 & 	 4311  & 	-20.88	 & 	 0.44 	 & 	1  &  B   &  G, C, R 	&   			 \\
	0895 	 & 	Sc 	 & 	 11.88	 & 	 2288  & 	-20.67	 & 	 0.50 	 & 	5  &  SD  &  G, B, C 	&   			 \\
	0925 	 & 	Scd  & 	 9.78	 & 	 554  & 	-20.17	 & 	 0.25 	 & 	5  &  B   &  G, C 		&   			 \\
	0977 	 & 	Sa 	 & 	 13.85	 & 	 4611  & 	-20.22	 & 	 0.43 	 & 	4  &  B   &  G, C 		&   			 \\
	0986 	 & 	Sab  & 	 11.44	 & 	 1984  & 	-20.62	 & 	 0.09 	 & 	2  &  B   &  G, C 		&  			 \\
	1068 	 & 	Sb 	 & 	 9.53	 & 	 1138  & 	-21.50	 & 	 0.04 	 & 	4  &  G   &  B 			&  ARP037, Grp 	 \\
	1073 	 & 	SBc  & 	 11.16	 & 	 1208  & 	-20.02	 & 	 0.37 	 & 	2  &  B   &  SD, C 		&   			 \\
	1187 	 & 	Sc 	 & 	 11.03	 & 	 1391  & 	-20.22	 & 	 0.33 	 & 	4  &  B   &  G, C, R 	&  Grp 			 \\
	1300 	 & 	Sbc  & 	 10.60	 & 	 1578  & 	-20.97	 & 	 0.15 	 & 	7  &  B   &  G, C, R 	&   			 \\
	1365 	 & 	Sb 	 & 	 9.83	 & 	 1638  & 	-21.75	 & 	 0.27 	 & 	8  &  G   &  B 			&  VV825 		 \\
	1385 	 & 	Sc 	 & 	 11.01	 & 	 1497  & 	-20.45	 & 	 0.17 	 & 	3  &  B   &  G 			&   			 \\
	1512 	 & 	Sa 	 & 	 10.75	 & 	 898  & 	-19.20	 & 	 0.82 	 & 	8  &  G   &    &  Grp 			 \\
	1566 	 & 	SABb & 	 9.98	 & 	 1502  & 	-21.27	 & 	 0.32 	 & 	6  &  G   &  B &   			 \\
	1667 	 & 	SABc & 	 12.22	 & 	 4567  & 	-21.83	 & 	 0.09 	 & 	2  &  SD  &  B &   			 \\
	1832 	 & 	Sbc  & 	 10.66	 & 	 1938  & 	-21.41	 & 	 0.15 	 & 	2  &  B   &  C &   			 \\
	2336 	 & 	Sbc  & 	 10.63	 & 	 2202  & 	-22.14	 & 	 0.17 	 & 	6  &  B   &  G &   			 \\
	2442 	 & 	Sbc  & 	 10.27	 & 	 1458  & 	-20.91	 & 	 0.32 	 & 	5  &  B   &  C &  Grp	 		 \\
	2460 	 & 	Sab  & 	 12.14	 & 	 1442  & 	-19.78	 & 	 0.46 	 & 	3  &  B   &   &   			 \\
	2528 	 & 	SABb & 	 13.35	 & 	 3918  & 	-20.50	 & 	 0.14 	 & 	3  &  SD  &  B &   			 \\
	2532 	 & 	SABc & 	 12.65	 & 	 5248  & 	-21.80	 & 	 0.29 	 & 	5  &  SD  &  B &   			 \\
	2771 	 & 	Sab  & 	 13.70	 & 	 5097  & 	-20.74	 & 	 0.79 	 & 	7  &  SD  &  G, B &   			 \\
	2861 	 & 	SBbc & 	 13.60	 & 	 5076  & 	-20.74	 & 	 0.17 	 & 	4  &  SD  &  B, C &   			 \\
	2989 	 & 	SABb & 	 12.99	 & 	 4143  & 	-20.86	 & 	 0.43 	 & 	7  &  HST &  B &  			 \\
	3261 	 & 	Sbc  & 	 11.35	 & 	 2564  & 	-21.34	 & 	 0.35 	 & 	6  &  B   &  C &   			 \\
	3319 	 & 	SBc  & 	 11.15	 & 	 742  & 	-19.54	 & 	 0.24 	 & 	4  &  SD  &  B, C &   		 	 \\
	3359 	 & 	Sc   & 	 10.75	 & 	 1013  & 	-20.57	 & 	 0.36 	 & 	5  &  B   &  G, SD &   			 \\
	3408 	 & 	Sc   & 	 14.07	 & 	 9507  & 	-21.70	 & 			 & 	3  &  SD  &  B, C, R &   			 \\
	3507 	 & 	SBb  & 	 11.86	 & 	 975  & 	-19.09	 & 	 0.18 	 & 	5  &  SD  &  B, C &   			 \\
	3513 	 & 	SBc  & 	 11.09	 & 	 1198  & 	-19.94	 & 	 0.14 	 & 	6  &  B   &  G, C &   			 \\
	3601 	 & 	SBab & 	 13.69	 & 	 8119  & 	-21.68	 & 	 0.17 	 & 	4  &  SD  &  &   			 \\
	3686 	 & 	SBbc & 	 11.69	 & 	 1157  & 	-19.63	 & 	 0.11 	 & 	5  &  B   &  SD, G &   			 \\
	3719 	 & 	Sbc  & 	 13.27	 & 	 5863  & 	-21.40	 & 	 0.34 	 & 	8  &  SD  &  B &   			 \\
	3726 	 & 	Sc   & 	 10.31	 & 	 864  & 	-20.72	 & 	 0.12 	 & 	5  &  B   &  SD, G, C &   			 \\
	3893 	 & 	SABc & 	 10.26	 & 	 962  & 	-21.00	 & 	 0.10 	 & 	3  &  B   &  SD, G, C &  M 			 \\
	3978 	 & 	SABb & 	 13.14	 & 	 9950  & 	-22.73	 & 	 0.14 	 & 	2  &  SD  &  B &   			 \\
	3992 	 & 	Sbc  & 	 10.08	 & 	 1047  & 	-21.31	 & 	 0.09 	 & 	2  &  B   &  SD &   			 \\
	4029 	 & 	SABb & 	 14.02	 & 	 6197  & 	-20.77	 & 	 0.21 	 & 	2  &  B   &  SD &   			 \\
	4030 	 & 	Sbc  & 	 10.86	 & 	 1463  & 	-20.84	 & 	 0.20 	 & 	5  &  SD  &  G, B &   			 \\
	4035 	 & 	SABb & 	 13.73	 & 	 1569  & 	-18.03	 & 	 0.84 	 & 	4  &  B   &  &   			 \\
	4079 	 & 	SABb & 	 13.27	 & 	 6086  & 	-21.47	 & 	 0.21 	 & 	8  &  SD  &  B &       		 \\
	4123 	 & 	Sc 	 & 	 11.60	 & 	 1327  & 	-19.91	 & 	 0.38 	 & 	2  &  B   &  SD, C &   			 \\
	4136 	 & 	Sc 	 & 	 11.92	 & 	 590  & 	-18.38	 & 	 0.27 	 & 	2  &  B   &  SD, C &   			 \\
	4141 	 & 	SBc  & 	 14.35	 & 	 1900  & 	-18.16	 & 	 0.71 	 & 	2  &  SD  &  B, C, G &   			 \\
	4145 	 & 	Scd  & 	 11.06	 & 	 1011  & 	-20.19	 & 	 0.19 	 & 	2  &  B   &  SD, G &   			 \\
	4151 	 & 	SABa & 	 11.09	 & 	 988  & 	-20.16	 & 	 0.15 	 & 	2  &  B   &  &   			 \\
	4156 	 & 	Sb 	 & 	 13.71	 & 	 6755  & 	-21.32	 & 	 1.22 	 & 	4  &  SD  &  B &   			 \\
	4210 	 & 	Sb 	 & 	 13.20	 & 	 2713  & 	-19.99	 & 	 0.14 	 & 	4  &  SD  &  B, C &   			 \\
	4254 	 & 	Sc 	 & 	 10.18	 & 	 2408  & 	-22.62	 & 	 0.19 	 & 	6  &  B   &  SD, G, 2M &   			 \\
	4304 	 & 	Sbc  & 	 12.14	 & 	 2623  & 	-20.66	 & 	 0.45 	 & 	5  &  B   &  C &   			 \\
	4319 	 & 	SBab & 	 12.21	 & 	 1443  & 	-19.81	 & 			 & 	1  &  B   &  &   			 \\
	4444 	 & 	SABb & 	 12.51	 & 	 2915  & 	-20.50	 & 	 0.48 	 & 	4  &  C   &  B &   			 \\
	4450 	 & 	Sab  & 	 10.48	 & 	 1955  & 	-21.90	 & 	 0.01 	 & 	4  &  G   &  SD, B &   			 \\
	4475 	 & 	SBbc & 	 14.01	 & 	 7388  & 	-21.20	 & 	 0.32 	 & 	7  &  SD  &  B &   			 \\
	4487 	 & 	Sc 	 & 	 11.27	 & 	 1036  & 	-19.67	 & 	 0.21 	 & 	2  &  C   &  SD, B &   			 \\
	4499 	 & 	SBbc & 	 13.24	 & 	 3353  & 	-20.09	 & 			 & 	3  &  B   &  C &   			 \\
	4536 	 & 	SABb & 	 10.32	 & 	 1807  & 	-21.85	 & 	 0.18 	 & 	7  &  G   &  B &   			 \\
	4579 	 & 	SABb & 	 10.12	 & 	 1517  & 	-21.74	 & 	 0.02 	 & 	3  &  B   &  SD &   			 \\
	4603 	 & 	SABc & 	 11.46	 & 	 2590  & 	-21.27	 & 	 0.25 	 & 	4  &  B   &  C &   			 \\
	4622 	 & 	Sa 	 & 	 13.22	 & 	 4468  & 	-20.77	 & 			 & 	6  &  B   &  C &   			 \\
	4653 	 & 	SABc & 	 12.61	 & 	 2624  & 	-20.33	 & 	 0.46 	 & 	5  &  B   &  SD, C &   			 \\
	4682 	 & 	SABc & 	 12.35	 & 	 2322  & 	-20.27	 & 	 0.19 	 & 	5  &  B   &  SD, C &   			 \\
	4701 	 & 	Sc 	 & 	 12.45	 & 	 723  & 	-17.84	 & 	 0.67 	 & 	2  &  SD  &  B, C &   			 \\
	4734 	 & 	Sc 	 & 	 13.90	 & 	 7525  & 	-21.32	 & 	 0.32 	 & 	2  &  SD  &  B, C &   			 \\
	4736 	 & 	Sab  & 	 8.54	 & 	 314  & 	-20.98	 & 	 0.01 	 & 	6  &  G   &  SD &   			 \\
	4897 	 & 	Sbc  & 	 12.87	 & 	 2558  & 	-19.97	 & 	 0.87 	 & 	5  &  B   &  G, C &   			 \\
	4902 	 & 	Sb 	 & 	 11.51	 & 	 2631  & 	-21.40	 & 	 0.21 	 & 	7  &  B   &  G, C &   			 \\
	4947 	 & 	Sb 	 & 	 11.97	 & 	 2405  & 	-20.65	 & 	 0.18 	 & 	6  &  B   &  C &   			 \\
	4965 	 & 	SABc & 	 12.23	 & 	 2264  & 	-20.27	 & 	 0.30 	 & 	8  &  B   &  C &   			 \\
	4981 	 & 	Sbc  & 	 11.65	 & 	 1678  & 	-20.31	 & 	 0.27 	 & 	4  &  B   &  G &   			 \\
	5020 	 & 	SABb & 	 12.94	 & 	 3362  & 	-20.57	 & 	 0.68 	 & 	5  &  SD  &  G, B &   			 \\
	5033 	 & 	Sc 	 & 	 10.08	 & 	 876  & 	-20.95	 & 	 0.19 	 & 	4  &  B   &  SD, G, C &   			 \\
	5101 	 & 	S0-a & 	 11.22	 & 	 1858  & 	-20.84	 & 	 0.12 	 & 	4  &  B   &   &   			 \\
	5112 	 & 	SBc  & 	 12.21	 & 	 972  & 	-19.04	 & 	 0.39 	 & 	2  &  SD  &  G, B, C &   			 \\
	5156 	 & 	SBb  & 	 11.74	 & 	 2986  & 	-21.33	 & 	 0.22 	 & 	6  &  B   &  2M &   			 \\
	5213 	 & 	Sb 	 & 	 14.48	 & 	 6884  & 	-20.55	 & 	 0.41 	 & 	2  &  SD  &  B, C, R &  VV018, M 	 \\
	5227 	 & 	Sb 	 & 	 13.60	 & 	 5235  & 	-20.84	 & 	 0.41 	 & 	5  &  SD  &  B, C &   			 \\
	5327 	 & 	Sb 	 & 	 13.15	 & 	 4354  & 	-20.89	 & 	 0.33 	 & 	4  &  SD  &  B &   			 \\
	5334 	 & 	Sc 	 & 	 12.52	 & 	 1380  & 	-19.12	 & 	 0.46 	 & 	3  &  B   &  SD, G, C &   			 \\
	5345 	 & 	Sa 	 & 	 13.34	 & 	 7255  & 	-21.79	 & 	 0.10 	 & 	3  &  SD  &  B &   			 \\
	5351 	 & 	SBb  & 	 12.57	 & 	 3613  & 	-21.16	 & 	 0.38 	 & 	6  &  B   &  SD, C, 2M &   	asym		 \\
	5395 	 & 	SABb & 	 11.72	 & 	 3468  & 	-21.92	 & 	 0.28 	 & 	5  &  B   &  SD, G &  VV048, ARP084, M  \\
	5494 	 & 	Sc 	 & 	 12.91	 & 	 2619  & 	-19.93	 & 	 0.61 	 & 	6  &  B   &  G, &   			 \\
	5584 	 & 	SABc & 	 12.33	 & 	 1648  & 	-19.69	 & 	 0.43 	 & 	3  &  B   &  SD &   			 \\
	5618 	 & 	SBc  & 	 13.40	 & 	 7140  & 	-21.71	 & 	 0.34 	 & 	3  &  SD  &  B, C &   			 \\
	5643 	 & 	Sc 	 & 	 9.95	 & 	 1190  & 	-21.00	 & 	 0.10 	 & 	5  &  B   &  C &   			 \\
	5653 	 & 	Sb 	 & 	 12.69	 & 	 3564  & 	-20.99	 & 	 0.12 	 & 	6  &  HST &  SD, 2M &   			 \\
	5655 	 & 		 & 	 16.71	 & 					 & 			 & 	 	 	 & 	4  &  B   &  SD &   			 \\
	5669 	 & 	SABc & 	 12.41	 & 	 1373  & 	-19.28	 & 	 0.57 	 & 	3  &  SD  &  B, C &  			 \\
	5674 	 & 	SABc & 	 13.06	 & 	 7473  & 	-22.16	 & 	 0.15 	 & 	3  &  B   &  SD, C &   			 \\
	5754 	 & 	SBb  & 	 13.74	 & 	 4404  & 	-20.41	 & 	 0.32 	 & 	11 &  SD  &  B  &  ARP297, M 	 \\
	5774 	 & 	SABc & 	 12.54	 & 	 1566  & 	-19.37	 & 	 0.61 	 & 	1  &  SD  &  B, G  &  M 			 \\
	5786 	 & 	Sbc  & 			 & 	 2981  & 			 & 			 & 	3  &  B   &  C &   			 \\
	5850 	 & 	Sb 	 & 	 11.40	 & 	 2547  & 	-21.50	 & 	 0.07 	 & 	8  &  G   &  SD, B &   			 \\
	5861 	 & 	SABc & 	 11.19	 & 	 1859  & 	-21.02	 & 	 0.18 	 & 	8  &  B   &  C &   			 \\
	5905 	 & 	Sb 	 & 	 12.98	 & 	 3391  & 	-20.66	 & 	 1.07 	 & 	6  &  SD  &  G, B, C, R &   			 \\
	5968 	 & 	SABb & 	 12.50	 & 	 5460  & 	-21.97	 & 	 0.21 	 & 	10 &  B   &  C, 2M &   			 \\
	6001 	 & 	Sc 	 & 	 14.07	 & 	 9974  & 	-21.79	 & 	 0.44 	 & 	5  &  SD  &  B, C &   			 \\
	6008 	 & 	Sb 	 & 	 13.72	 & 	 4861  & 	-20.60	 & 	 0.45 	 & 	4  &  SD  &  B, C &   			 \\
	6035 	 & 	Sc 	 & 	 13.73	 & 	 4756  & 	-20.55	 & 	 0.24 	 & 	4  &  SD  &  B, C, 2M &   			 \\
	6217 	 & 	Sbc  & 	 11.46	 & 	 1368  & 	-20.45	 & 	 0.27 	 & 	5  &  SD  &  G, B, C &  ARP185 		 \\
	6221 	 & 	Sc 	 & 	 9.69	 & 	 1485  & 	-21.70	 & 	 0.21 	 & 	4  &  B   &  C &   			 \\
	6267 	 & 	Sc 	 & 	 13.29	 & 	 2980  & 	-20.04	 & 	 0.23 	 & 	3  &  B   &  B, C &   			 \\
	6384 	 & 	SABb & 	 10.55	 & 	 1664  & 	-21.51	 & 	 0.19 	 & 	4  &  B   &  SD, C &   			 \\
	6484 	 & 	Sb 	 & 	 12.73	 & 	 3114  & 	-20.68	 & 	 0.66 	 & 	3  &  B   &  SD, C &   			 \\
	6691 	 & 	Sbc  & 	 13.27	 & 	 5883  & 	-21.50	 & 	 0.20 	 & 	2  &  B   &  C, 2M &   			 \\
	6699 	 & 	SABb & 	 12.39	 & 	 3391  & 	-20.94	 & 	 0.07 	 & 	7  &  B   &  C &   			 \\
	6744 	 & 	Sbc  & 	 8.61	 & 	 851  & 	-21.33	 & 	 1.09 	 & 	8  &  G   &  B, C &  G, UNGC 	 \\
	6753 	 & 	Sb 	 & 	 11.51	 & 	 3176  & 	-21.68	 & 	 0.17 	 & 	4  &  B   &  G, C &   			 \\
	6769 	 & 	SABb & 	 12.10	 & 	 3813  & 	-21.49	 & 			 & 	6  &  B   &  C &  VV304, M 	 \\
	6770 	 & 	Sb 	 & 	 12.24	 & 	 3842  & 	-21.37	 & 	 0.33 	 & 	3  &  B   &  G, C &  VV304, M 	 \\
	6780 	 & 	SABc & 	 12.90	 & 	 3493  & 	-20.51	 & 	 0.25 	 & 	2  &  B   &  C &   			 \\
	6782 	 & 	Sa 	 & 	 12.18	 & 	 3923  & 	-21.48	 & 	 0.11 	 & 	6  &  G   &  B &   			 \\
	6845 	 & 	SBbc & 	 13.19	 & 	 6679  & 	-21.69	 & 			 & 	3  &  B   &   &   			 \\
	6878 	 & 	SABb & 	 13.20	 & 	 5853  & 	-21.40	 & 			 & 	6  &  B   &  C &   			 \\
	6919 	 & 	SABc & 	 13.26	 & 	 6727  & 	-21.65	 & 			 & 	2  &  C   &  B, C &   			 \\
	6923 	 & 	SBb  & 	 11.89	 & 	 2831  & 	-21.11	 & 	 0.46 	 & 	4  &  B   &  C &   			 \\
	6935 	 & 	SABa & 	 12.56	 & 	 4589  & 	-21.48	 & 			 & 	4  &  G   &  C &   			 \\
	6941 	 & 	Sb 	 & 	 13.48	 & 	 6219  & 	-21.32	 & 	 0.43 	 & 	2  &  C   &  SD, B  &   			 \\
	6943 	 & 	Sc 	 & 	 11.45	 & 	 3116  & 	-21.65	 & 	 0.43 	 & 	6  &  C   &  G, B &   			 \\
	6949 	 & 	Sc 	 & 	 12.76	 & 	 2763  & 	-20.48	 & 	 0.33 	 & 	5  &  C   &  B &   			 \\
	6951 	 & 	SABb & 	 10.05	 & 	 1425  & 	-21.92	 & 	 0.04 	 & 	4  &  C   &  B &   			 \\
	6955 	 & 	SABb & 	 14.35	 & 	 8168  & 	-21.05	 & 			 & 	5  &  SD  &  B, C &   			 \\
	6962 	 & 	SABa & 	 12.37	 & 	 4202  & 	-21.58	 & 	 0.33 	 & 	5  &  G   &  SD, B, R &  M 			 \\
	6976 	 & 	SABb & 	 14.35	 & 	 5998  & 	-20.36	 & 			 & 	6  &  SD  &  B &   			 \\
	6984 	 & 	Sc 	 & 	 12.66	 & 	 4663  & 	-21.42	 & 	 0.40 	 & 	6  &  C   &  B &   			 \\
	7038 	 & 	SABc & 	 11.94	 & 	 4932  & 	-22.25	 & 	 0.34 	 & 	8  &  B   &  C &   			 \\
	7059 	 & 	SABc & 	 11.56	 & 	 1734  & 	-20.19	 & 	 0.70 	 & 	2  &  C   &  B &   			 \\
	7065A 	 & 	SABc & 	 14.46	 & 	 7404  & 	-20.70	 & 			 & 	3  &  SD  &  B &  			 \\
	7070 	 & 	Sc 	 & 	 12.54	 & 	 2397  & 	-20.00	 & 	 0.82 	 & 	4  &  B   &  C &   			 \\
	7083 	 & 	SBbc & 	 11.30	 & 	 3106  & 	-21.80	 & 	 0.36 	 & 	7  &  B   &  C &   			 \\
	7102 	 & 	Sb 	 & 	 13.47	 & 	 4846  & 	-20.79	 & 	 0.41 	 & 	4  &  B   &  C &   			 \\
	7125 	 & 	SABc & 	 12.31	 & 	 3078  & 	-20.79	 & 	 1.68 	 & 	10 &  B   &  C &   			 \\
	7171 	 & 	SBb  & 	 12.43	 & 	 2717  & 	-20.55	 & 	 0.41 	 & 	3  &  B   &  C &  			 \\
	7221 	 & 	Sbc  & 	 12.41	 & 	 4356  & 	-21.54	 & 	 0.30 	 & 	3  &  C   &  B, G  &   			 \\
	7252 	 & 	S0 	 & 	 12.59	 & 	 4724  & 	-21.54	 & 	 0.04 	 & 	5  &  HST &   &  ARP226, M 	 \\
	7257 	 & 	SABb & 	 12.95	 & 	 4902  & 	-21.31	 & 	 0.20 	 & 	4  &  SD  &  B, C &   			 \\
	7309 	 & 	SABc & 	 12.70	 & 	 4006  & 	-21.10	 & 	 0.19 	 & 	5  &  B   &  C &   			 \\
	7323 	 & 	Sb 	 & 	 13.62	 & 	 5601  & 	-20.98	 & 	 0.35 	 & 	2  &  SD  &   &  M 			 \\
	7329 	 & 	SBbc & 	 12.04	 & 	 3251  & 	-21.15	 & 	 0.46 	 & 	6  &  C   &  B, G &   			 \\
	7412 	 & 	SBb  & 	 11.51	 & 	 1710  & 	-20.25	 & 	 0.25 	 & 	4  &  C   &  B, G &   			 \\
	7418 	 & 	Sc 	 & 	 11.42	 & 	 1447  & 	-19.97	 & 	 0.27 	 & 	5  &  B   &  C, B &   			 \\
	7421 	 & 	Sbc  & 	 12.41	 & 	 1801  & 	-19.51	 & 	 0.15 	 & 	5  &  B   &  C, G &   			 \\
	7424 	 & 	Sc 	 & 	 10.52	 & 	 937  & 	-19.78	 & 	 0.85 	 & 	5  &  B   &  C, G &   			 \\
	7535 	 & 	Scd  & 	 13.95	 & 	 4604  & 	-20.20	 & 	 0.65 	 & 	2  &  SD  &  B, C &   			 \\
	7678 	 & 	Sc 	 & 	 12.04	 & 	 3488  & 	-21.55	 & 	 0.16 	 & 	6  &  B   &  C, G, SD, 2M  &  VV359, ARP028  \\
	7691 	 & 	Sbc  & 	 13.58	 & 	 4038  & 	-20.31	 & 	 0.36 	 & 	4  &  SD  &   &   			 \\
	7713A 	 & 	Sc 	 & 	 12.88	 & 	 3008  & 	-20.19	 & 	 0.27 	 & 	5  &  C   &  B &   			 \\
	7714 	 & 	Sb 	 & 	 12.53	 & 	 2797  & 	-20.51	 & 	 0.42 	 & 	2  &  SD  &  B, C  &  VV051, ARP284, M  \\
	\hline
	IC		 &  			 &      	 &        		 &         &  		 &    &  	  & 	 &   		 \\
	\hline
	0004     &  Sc    	 & 13.69	 & 5011  & -20.65	 & 0.17	 & 2  &  SD  &  B	 &   		 \\
	0167     &  Sc    	 & 12.92	 & 2934  & -19.39	 & 0.35	 & 4  &  G  &  SD &  ARP031  \\
	0173     &  SBbc  	 & 14.57	 & 13965  & -21.99	 & 0.24	 & 6  &  SD  &  B	 &   		 \\
	0209     &  SBbc  	 & 13.35	 & 3964  & -20.41	 & 0.17	 & 2  &  SD  & 	 &   		 \\
	0221     &  Sc    	 & 13.13	 & 5088  & -21.25	 & 0.46	 & 4  &  C  &  B	 &   		 \\
	0342     &  SABc  	 & 6.14	 & 25  & -21.54	 & 0.47	 & 7  &  B  &  C, R	 &   		 \\
	0370     &  Sc    	 & 14.39	 & 9727  & -21.34	 & 0.52	 & 4  &  B  &  C	 &   		 \\
	0382     &  SABc  	 & 12.74	 & 4998  & -21.5	 & 0.53	 & 4  &  B  & 	C, G &   		 \\
	0438     &  SABc  	 & 12.43	 & 3123  & -20.56	 & 0.58	 & 4  &  C  &  B, G	 &   		 \\
	0492     &  SBbc  	 & 13.9	 & 5148  & -20.51	 & 0.35	 & 5  &  SD  & 	 &   		 \\
	0498     &  Sb    	 & 14.27	 & 10139  & -21.59	 & 0.38	 & 2  &  SD  & 	 &  VV526 	 \\
	0503     &  Sa    	 & 14		 & 4125  & -19.87	 & 1.39	 & 3  &  SD  & 	 &   		 \\
	0509     &  SABc  	 & 13.56	 & 5490  & -20.98	 & 0.45	 & 5  &  B   &  SD 	 &   		 \\
	0512     &  SABc  	 & 12.7	 & 1614  & -19.5	 & 0.40	 & 6  &  G   &  B	 &   		 \\
	0527     &  SBc   	 & 14.43	 & 6868  & -20.61	 & 0.86	 & 7  &  SD  &  B	 &   		 \\
	0539     &  Sc    	 & 13.97	 & 7043  & -21.07	 & 0.29	 & 5  &  SD  &  B	 &   		 \\
	0577     &  Sc    	 & 14.65	 & 9014  & -20.95	 & 0.33	 & 6  &  SD  &  B	 &   		 \\
	0616     &  Sc    	 & 14.27	 & 5776  & -20.38	 & 0.42	 & 5  &  SD  &  B	 &  		 \\
	0651     &  SBd   	 & 12.9	 & 4491  & -21.17	 & 0.28	 & 3  &  SD  &  		 &   		 \\
	0900     &  SABc  	 & 13.3	 & 7066  & -21.8	 & 0.35	 & 4  &  SD  &  C, B	 &   		 \\
	0983     &  SBbc  	 & 12.49	 & 5442  & -22.06	 & 0.67	 & 5  &  SD  &  C, B	 &  ARP117  \\
	0992     &  SABb  	 & 14.46	 & 7782  & -20.84	 & 0.64	 & 2  &  SD  &  B	 &   		 \\
	1093     &  SABb  	 & 14.5	 & 13370  & -21.99	 & 0.47	 & 4  &  SD  &  B, G	 &   		 \\
	1132     &  Sc    	 & 14.07	 & 4524  & -20.1	 & 0.65	 & 5  &  SD  &  C, B	 &   		 \\
	1142     &  SBc   	 & 14.51	 & 13973  & -22.09	 & 0.70	 & 4  &  SD  &  C, B, 2M	 &  M 		 \\
	1149     &  Sbc   	 & 13.76	 & 4681  & -20.47	 & 0.41	 & 4  &  SD  &  C, B	 &  M 		 \\
	1236     &  Sc    	 & 13.79	 & 6026  & -20.99	 & 0.29	 & 3  &  SD  &  C, B	 &  VV442, M  \\
	1269     &  Sbc   	 & 12.7	 & 6115  & -22.11	 & 0.26	 & 6  &  B   &  C	 &   		 \\
	1301     &  Sc    	 & 14.1	 & 3990  & -19.86	 & 1.36	 & 3  &  C   &  B	 &   		 \\
	1377     &  SBab  	 & 14.22	 & 9039  & -21.4	 & 		 & 5  &  SD  & 	 &   		 \\
	1516     &  Sbc   	 & 13.83	 & 7278  & -21.29	 & 1.08	 & 7  &  SD  &  B	 &   		 \\
	1525     &  Sb    	 & 12.52	 & 5010  & -21.89	 & 0.31	 & 5  &  B   &  C	 &   		 \\
	1543     &  Sbc   	 & 13.94	 & 5595  & -20.65	 & 0.36	 & 2  &  SD  &  B	 &   		 \\
	1562     &  SBc   	 & 13.46	 & 3728  & -20.12	 & 0.65	 & 4  &  C   &  B	 &   		 \\
	1607     &  Sc    	 & 14.14	 & 5438  & -20.34	 & 0.66	 & 3  &  SD  & 	 &   		 \\
	1666     &  Sc    	 & 14.1	 & 4881  & -20.2	 & 0.38	 & 2  &  B   &  SD	 &   		 \\
	1734     &  Sc    	 & 13.29	 & 4924  & -20.9	 & 0.34	 & 3  &  B   &  	 &   		 \\
	1764     &  Sb    	 & 13.81	 & 5066  & -20.56	 & 0.45	 & 5  &  B   &  C	 &   		 \\
	1852     &  Sc    	 & 14		 & 8428  & -21.45	 & 		 & 4  &  B   &  	 &   		 \\
	1953     &  Sc    	 & 11.76	 & 1864  & -20.21	 & 0.15	 & 2  &  B   & 	 &   		 \\
	2226     &  SABa  	 & 13.88	 & 10876  & -22.14	 & 0.13	 & 5  &  SD  &  B	 &   		 \\
	2473     &  Sbc   	 & 14.32	 & 8070  & -21.07	 & 0.31	 & 3  &  G   &  	 &   		 \\
	2490     &  SABb  	 & 14.3	 & 7333  & -20.88	 & 0.97	 & 4  &  SD  & 	 &   		 \\
	2522     &  Sc    	 & 12.44	 & 3017  & -20.32	 & 0.61	 & 2  &  B   &  G, C	 &   		 \\
	2537     &  Sc    	 & 12.11	 & 2788  & -20.65	 & 0.34	 & 4  &  B   &  G	 &   		 \\
	2548     &  SBbc  	 & 13.18	 & 4409  & -20.77	 & 0.39	 & 6  &  B   &  C	 &   		 \\
	2556     &  Scd   	 & 13.37	 & 2505  & -19.46	 & 0.47	 & 2  &  B   &  C	 &   		 \\
	2580     &  Sc    	 & 12.93	 & 3140  & -20.26	 & 0.51	 & 6  &  B   &  C, R	 &   		 \\
	2582     &  Sbc   	 & 13.32	 & 4161  & -20.5	 & 0.21	 & 7  &  B   &  C, R	 &   		 \\
	2604     &  SBm   	 & 14.5	 & 1628  & -17.57	 & 1.75	 & 1  &  SD  &  B, C, R	 &  VV538, M  \\
	2627     &  SABc  	 & 11.89	 & 2088  & -20.41	 & 0.28	 & 5  &  B   &  G, C, 2M	 &   		 \\
	2947     &  SBm   	 & 14.3	 & 12705  & -22.08	 & 0.45	 & 3  &  SD  &  B	 &  asym	 \\
	2956     &  Sbc   	 & 14.62	 & 9056  & -21.02	 & 0.47	 & 5  &  B   &  SD  &   		 \\
	3062     &  Sc    	 & 14.32	 & 7860  & -21	 & 0.31	 & 5  &  SD  &  B	 &  		 \\
	3109     &  Sbc   	 & 14.46	 & 12853  & -21.93	 & 0.37	 & 4  &  SD  &  B	 &   		 \\
	3115     &  Sc    	 & 13.3	 & 734  & -18.64	 & 0.37	 & 6  &  B   &  G 	 &  VV431, Grp  \\
	3156     &  SBcd  	 & 14.33	 & 5874  & -20.37	 & 0.30	 & 2  &  B   &  SD	 &   		 \\
	3253     &  Sc    	 & 11.62	 & 2706  & -20.65	 & 0.24	 & 3  &  B   &  C	 &   		 \\
	3267     &  Sc    	 & 13.99	 & 1236  & -18.87	 & 0.20	 & 5  &  SD  &  B	 &   		 \\
	3271     &  SABc  	 & 14.42	 & 7211  & -20.71	 & 0.52	 & 3  &  SD  &  B, C	 &  M 		 \\
	3376     &  SBa   	 & 14.07	 & 7130  & -21.06	 & 0.65	 & 2  &  SD  &  B	 &   		 \\
	3407     &  SBb   	 & 14.14	 & 7004  & -20.95	 & 0.31	 & 3  &  B   &  SD	 &   		 \\
	3709     &  Sbc   	 & 14.68	 & 14311  & -21.95	 & 0.61	 & 4  &  SD  &  B	 &   		 \\
	3827     &  Sc    	 & 13.58	 & 4303  & -20.39	 & 0.29	 & 4  &  B   &  	 &   		 \\
	3829     &  SABa  	 & 13.26	 & 3564  & -20.25	 & 0.74	 & 4  &  B   &  C	 &   		 \\
	3896A    &  Scd   	 & 11.88	 & 2142  & -20.42	 & 0.13	 & 1  &  B	  &  C  &   		 \\
	4219     &  SBb   	 & 13.18	 & 3654  & -20.38	 & 0.14	 & 2  &  B   &  2M	 &   		 \\
	4229     &  Sb    	 & 14.16	 & 6953  & -20.89	 & 0.46	 & 2  &  SD  &  B, C	 &  M 		 \\
	4237     &  SBb   	 & 12.48	 & 2658  & -19.87	 & 0.24	 & 4  &  B   &  C	 &   		 \\
	4248     &  Sc    	 & 13.51	 & 4105  & -20.31	 & 0.74	 & 4  &  B   &  	 &   		 \\
	4270     &  Sbc   	 & 14.16	 & 7964  & -21.13	 & 0.70	 & 5  &  B   &  C	 &   		 \\
	4341     &  Sc    	 & 14.64	 & 2344  & -18.24	 & 1.96	 & 4  &  B   &  SD	 &   		 \\
	4359     &  Sc    	 & 13.16	 & 4130  & -20.64	 & 0.54	 & 3  &  B   &  C, 2M	 &   		 \\
	4366     &  Sc    	 & 12.75	 & 4615  & -21.32	 & 0.31	 & 5  &  B   &  C	 &   		 \\
	4367     &  SABc  	 & 12.54	 & 4054  & -21.24	 & 0.23	 & 6  &  B   &  C	 &   		 \\
	4388     &  Sbc   	 & 13.79	 & 4010  & -19.99	 & 0.59	 & 3  &  B   &  	 &   		 \\
	4397     &  Sbc   	 & 13.7	 & 4381  & -20.41	 & 0.52	 & 5  &  SD  &  B	 &   		 \\
	4441     &  Sc    	 & 13.65	 & 4452  & -20.34	 & 0.45	 & 2  &  B   &  C	 &  Grp 		 \\
	4444     &  SABb  	 & 11.2	 & 1958  & -20.91	 & 0.13	 & 5  &  B   &  2M	 &   		 \\
	4479     &  Sc    	 & 14.67	 & 13593  & -21.86	 & 0.64	 & 3  &  SD  &  B, C	 &   		 \\
	4538     &  SABc  	 & 12.13	 & 2862  & -20.94	 & 0.12	 & 5  &  B   &  C	 &   		 \\
	4567     &  Sc    	 & 13.24	 & 5728  & -21.31	 & 0.21	 & 4  &  SD  &  2M	 &  M 		 \\
	4585     &  SBb   	 & 12.05	 & 3645  & -21.46	 & 		 & 6  &  B   &  C	 &   		 \\
	4633     &  Sc    	 & 12.14	 & 2942  & -20.44	 & 0.56	 & 6  &  B   &  C	 &   		 \\
	4641     &  Sc    	 & 13.7	 & 5333  & -20.64	 & 0.16	 & 7  &  B   &  C	 &   		 \\
	4646     &  Sc    	 & 11.83	 & 3172  & -21.36	 & 0.53	 & 6  &  B   &  C, 2M	 &   		 \\
	4661     &  Sc    	 & 12.95	 & 4828  & -21.16	 & 0.61	 & 5  &  B   &  C	 &   		 \\
	4682     &  SBbc  	 & 12.38	 & 3570  & -21.05	 & 0.47	 & 4  &  B   &  C	 &   		 \\
	4688     &  Sc    	 & 13.71	 & 6026  & -21.05	 & 0.21	 & 3  &  B   &  SD	 &   		 \\
	4722     &  Sc    	 & 12.83	 & 4817  & -21.3	 & 		 & 6  &  B   &  	 &   		 \\
	4729     &  Sc    	 & 12.44	 & 4434  & -21.49	 & 0.29	 & 8  &  B   &  C	 &   		 \\
	4769     &  SBbc  	 & 13.26	 & 4535  & -20.73	 & 0.60	 & 5  &  B   &  C	 &   		 \\
	4836     &  SBc   	 & 12.82	 & 4604  & -21.21	 & 1.07	 & 5  &  B   &  G, 2M	 &   		 \\
	4839     &  Sbc   	 & 12.77	 & 2717  & -20.07	 & 0.50	 & 6  &  B   &  2M 	 &   		 \\
	4852     &  SBc   	 & 12.84	 & 4425  & -21.09	 & 0.36	 & 5  &  B   &  C	 &   		 \\
	4857     &  Sc    	 & 12.94	 & 4674  & -21.13	 & 0.38	 & 3  &  B   &  C	 &   		 \\
	4876     &  SABc  	 & 13.7	 & 5577  & -20.78	 & 0.63	 & 5  &  B   &  C	 &   		 \\
	4901     &  SABc  	 & 11.44	 & 2139  & -20.11	 & 0.37	 & 9  &  B   &  C	 &   		 \\
	4933     &  SBbc  	 & 12.4	 & 4910  & -21.79	 & 0.37	 & 4  &  B   &  G	 &   		 \\
	4998     &  SBc   	 & 13.42	 & 6662  & -21.48	 & 		 & 4  &  B	  &  C	 &   		 \\
	5005     &  SBc   	 & 13.06	 & 3094  & -20.16	 & 0.71	 & 6  &  C   &  B	 &   		 \\
	5092     &  SBc   	 & 12.74	 & 3246  & -20.48	 & 0.42	 & 9  &  B   &  	 &   		 \\
	5116     &  SBbc  	 & 13.6	 & 3791  & -19.97	 & 0.47	 & 3  &  B   &  C	 &   		 \\
	5141     &  Sbc   	 & 13.09	 & 4477  & -20.86	 & 0.64	 & 4  &  B   &  C	 &   		 \\
	5188     &  SABc  	 & 13.47	 & 4614  & -20.55	 & 0.47	 & 5  &  B   &  C	 &  M, Grp 	 \\
	5261     &  SBc   	 & 13.56	 & 3238  & -19.74	 & 0.46	 & 4  &  B   &  C	 &   		 \\
	5325     &  Sbc   	 & 11.71	 & 1507  & -19.75	 & 0.12	 & 8  &  C	  &  G, 2M  &   		 \\
	\hline
	\multicolumn{1}{p{13mm}}
	{\footnotesize
		\begin{tabular}{lrp{130mm}}
			$^*$ Abbreviations:	& C  & --- DSS colored;\\
				& B  & --- DSS2 Blue (XJ+S);\\%[-5pt]
				& R  & --- DSS2 Red (F+R);\\%[-5pt]
				& 2M  & --- 2MASS;\\
				& G  & --- GALEX;\\
				& SD & --- SDSS DR9 color;\\
				& HST& --- Hubble Space Telescope (optical).\\

		\end{tabular}}\\
		\multicolumn{2}{p{13mm}}
		{\footnotesize
			\begin{tabular}{lrp{130mm}}
				$^{**}$ Remarks: 	& VV 	& --- the number of the object in the Vorontsov-Velyaminov catalog.\\%[-5pt]
			& ARP 	& --- the number of the object in the catalog of peculiar galaxies.\\%[-5pt]
			& M		& --- the object belongs to a pair or a group of galaxies according to HyperLeda data.\\
			& Grp 	& --- the object is included in the catalog of Makarov and Karachentsev~\cite{Makarov_Karachentsev_2011_MNRAS} .\\
			& UNGC 	& --- the object is included into the catalog of Karachentsev et al.~\cite{Karachentsev_Makarov_Kaisina_2013_AJ} and has large tidal index ($\Theta\geqslant 2.0$).\\
			& asym 	& --- the object is asymmetric with pronounced signs of interaction. \\
			\end{tabular}}\\		
		\end{longtable*}

Note that unlike the sample constructed by Chernin
et al.~\cite{Chernin_etal_2001}, only 12 of our selected 276 galaxies are
mentioned in the Catalog of interacting galaxies of
Vorontsov-Velyaminov (VV). Several more galaxies
have spiral patterns with a characteristic perturbed
stricture or signs of strong interactions in
the past (e.g., NGC~0060, NGC~1068, NGC~2442, NGC~5774, IC~1142, IC~2956, IC~4441, IC~4567). When analyzing our sample we used,
in addition to the VV catalog, the catalog of peculiar
galaxies (ARP), tidal index estimates for Local group
galaxies~\cite{Makarov_Karachentsev_2011_MNRAS, Karachentsev_Makarov_Kaisina_2013_AJ}, and information from the HyperLeda
database, namely, the $multiple$ parameter, which is
equal to $M$ if the object belongs to a group. The small
fraction of interacting galaxies in our sample may be
due to the fact that objects with the most conspicuous
signs of interaction have already been included into
the catalog of A. D. Chernin et al.~\cite{Chernin_etal_2001}.
%, степень <<вовлеченности>> объекта в группу \cite{}
%ќчевидно, среди ближайших объектов основна€ часть взаимодействующих галактик с вереницами уже попала в каталог~\cite{Chernin_etal_2001}.
%“аким образом, в нашей выборке менее 9\% галактик €вл€ютс€ взаимодействующими. ѕоэтому мы не будем рассматривать отдельно этот класс объектов.

\subsection{Statistical Properties}

To analyze the possibility of merging our sample
with that of catalog~\cite{Chernin_etal_2001}, we identified УrowsФ in six
randomly selected objects from~\cite{Chernin_etal_2001} using the corresponding
SDSS/DSS images instead of the Palomar
Atlas images and found our results to differ appreciably
from those of Chernin et al. in terms of the
number of УrowsФ (Table \ref{tab:num_row_sravnenie})  for all objects except for
NGC~1637. The discrepancy is most significant in the
case of NGC~5921, and this can be explained by the
fact that modern images of the central region of the
galaxy have higher resolution unattainable with Palomar
Atlas images. In view of this result, which is due
to the large difference between the quality of initial
observational data, we consider it unfeasible to merge
the samples in order to analyze the distributions of the
number of УrowsФ, their linear and angular sizes.
\begin{table}
	\setcaptionmargin{0mm} \captionstyle{normal}
	\caption{Number of УrowsФ in randomly selected galaxies
from the catalog of Chernin et al.~\cite{Chernin_etal_2001},found by inspecting
Palomar Atlas and SDSS/DSS images}
	\label{tab:num_row_sravnenie}
	\medskip
	\begin{tabular}{|c|c|c|}
		\hline
		Galaxy & Number  & Our estimate  \\
        & of УrowsФ according & of the number \\
		& to~\cite{Chernin_etal_2001}& of УrowsФ\\
		\hline
		NGC~0514 & 2 & 4  \\
		NGC~1637 & 3 & 3  \\
		NGC~5597 & 2 & 4  \\
		NGC~5921 & 3 & 8  \\
		NGC~7229 & 2 & 3  \\
		NGC~7755 & 4 & 5  \\
		\hline
	\end{tabular}
\end{table}

At the same time, the distributions of other properties
of galaxies with polygonal structures from~\cite{Chernin_etal_2001}, agree well with the corresponding distributions reported
in this paper, and we show them both separately
(for the sample from~\cite{Chernin_etal_2001} with the updated parameter
values from HyperLeda database, and for our
sample) and for the combined sample.

Let us now compare the statistical properties of
our sample with the results of Chernin et al.~\cite{Chernin_etal_2001}. Figure
2 shows the distributions of morphological types.
This distribution for our galaxy is slightly shifted
toward earlier-type galaxies compared to the corresponding
distribution in~\cite{Chernin_etal_2001}.

% —оставной рисунок из двух eps-файлов. ¬ставлен на две колонки текста.

% Fig 2
\begin{figure*}
  \setcaptionmargin{5mm} \onelinecaptionstrue \captionstyle{normal}
  \begin{minipage}[h]{0.325\linewidth}
  	\center{\includegraphics[width=1\linewidth]{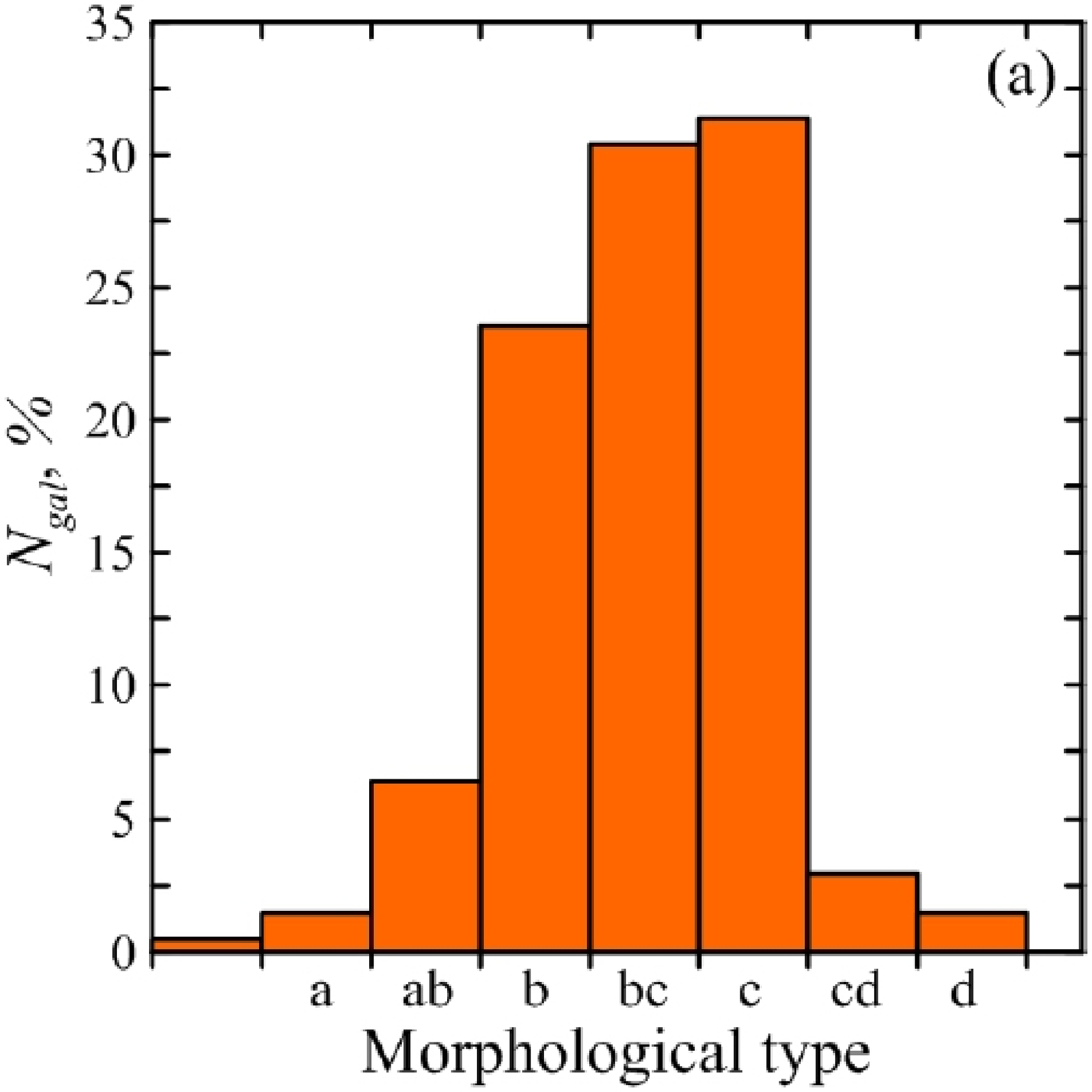} \\ } %{N_Mtipe_new_Chernin.jpg} \\ а)}
  \end{minipage}
  \hfill
  \begin{minipage}[h]{0.325\linewidth}
  	\center{\includegraphics[width=1\linewidth]{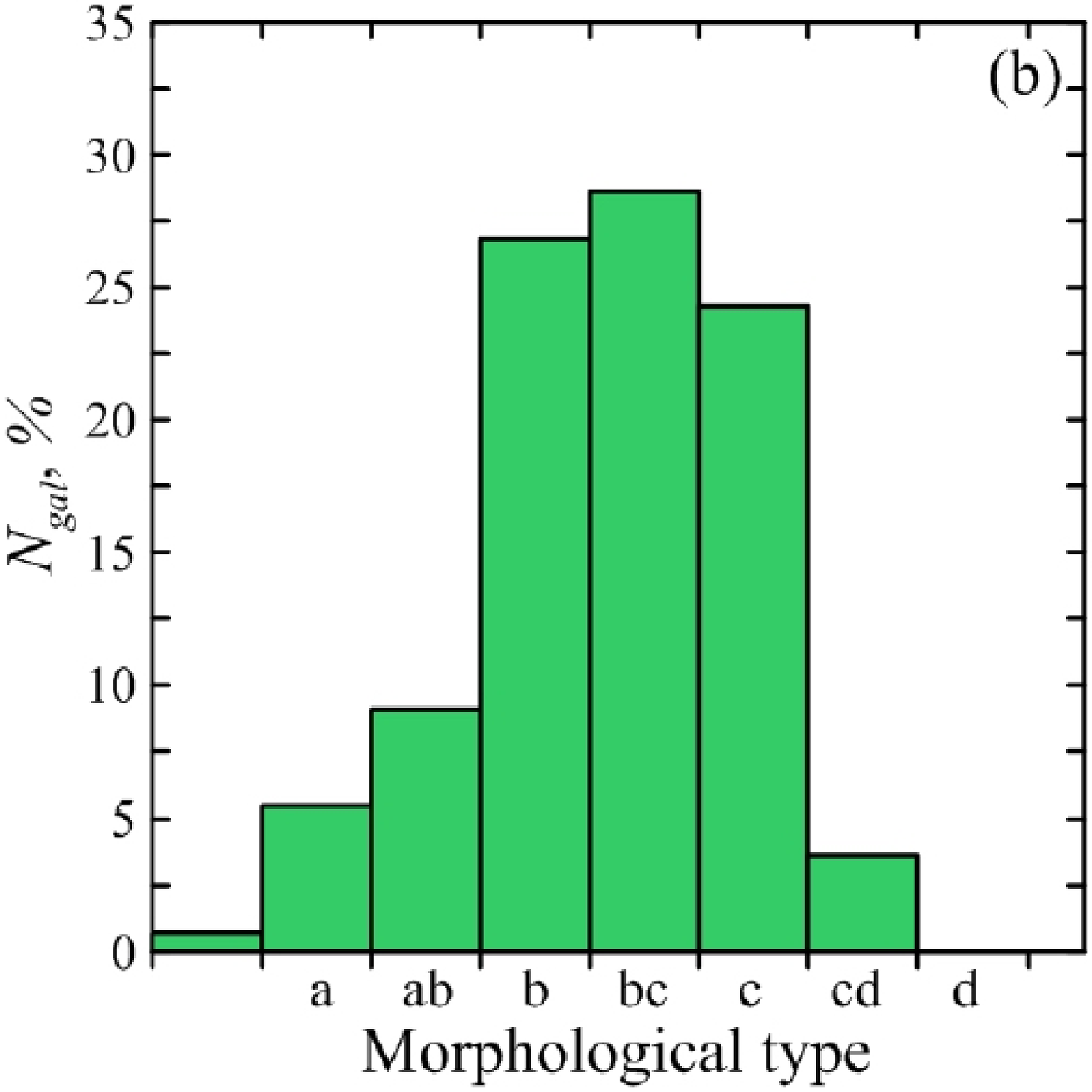} \\ }
  \end{minipage}
  \hfill
  \begin{minipage}[h]{0.325\linewidth}
  	\center{\includegraphics[width=1\linewidth]{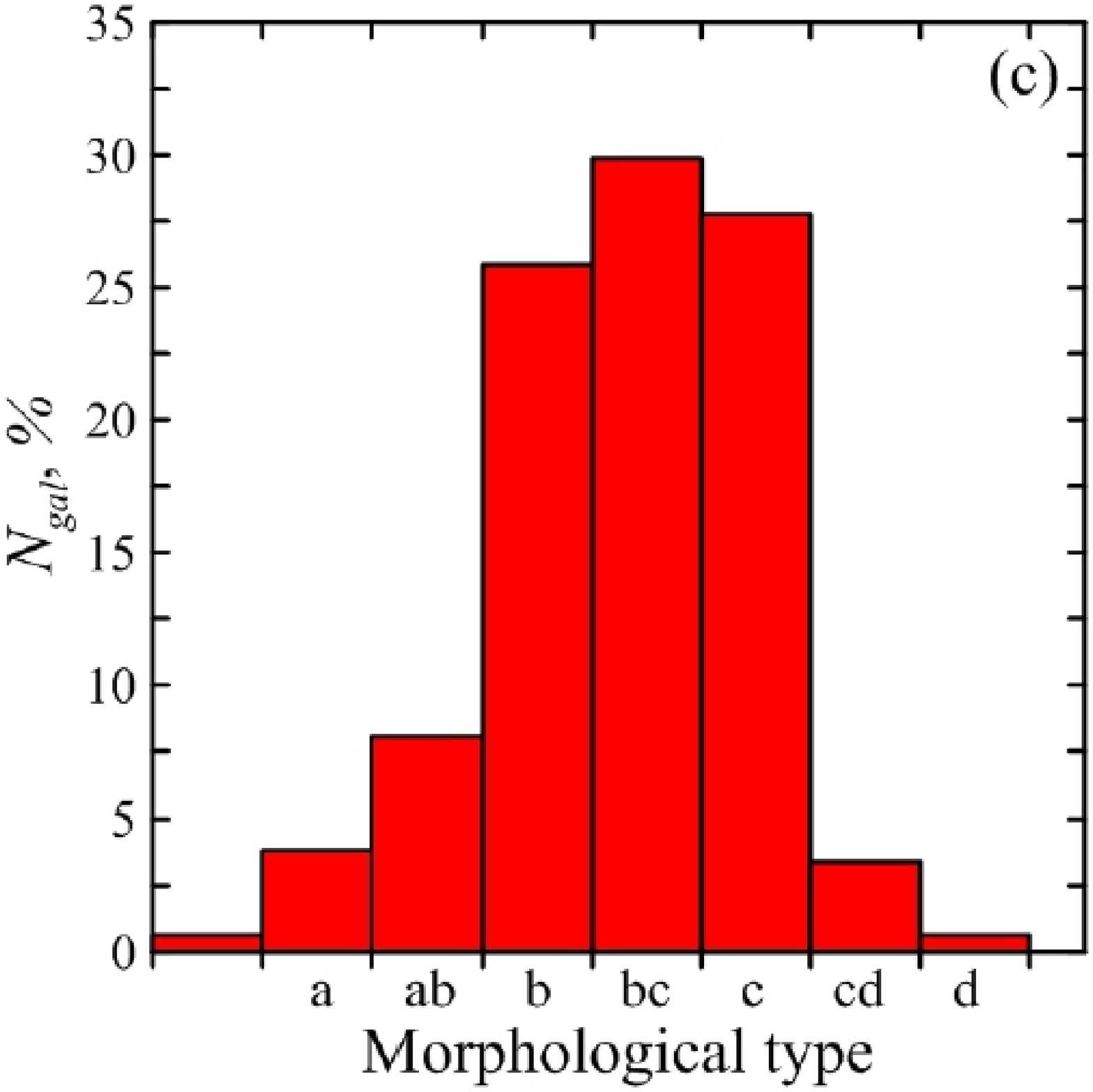} \\ }
  \end{minipage}
  \caption{Distribution ofmorphological types of galaxies: (a)
for the sample of Chernin et al.~\cite{Chernin_etal_2001} updated with information
from the HyperLeda database; (b) for our sample of
galaxies; (c) combined histogram for the merged sample.} %% подпись к рисунку
  	\label{fig2:MTipe}
\end{figure*}

\begin{figure*}
	\setcaptionmargin{5mm} \onelinecaptionstrue \captionstyle{normal}
		\begin{minipage}[h]{0.49\linewidth}
			\center{\includegraphics[width=1\linewidth]{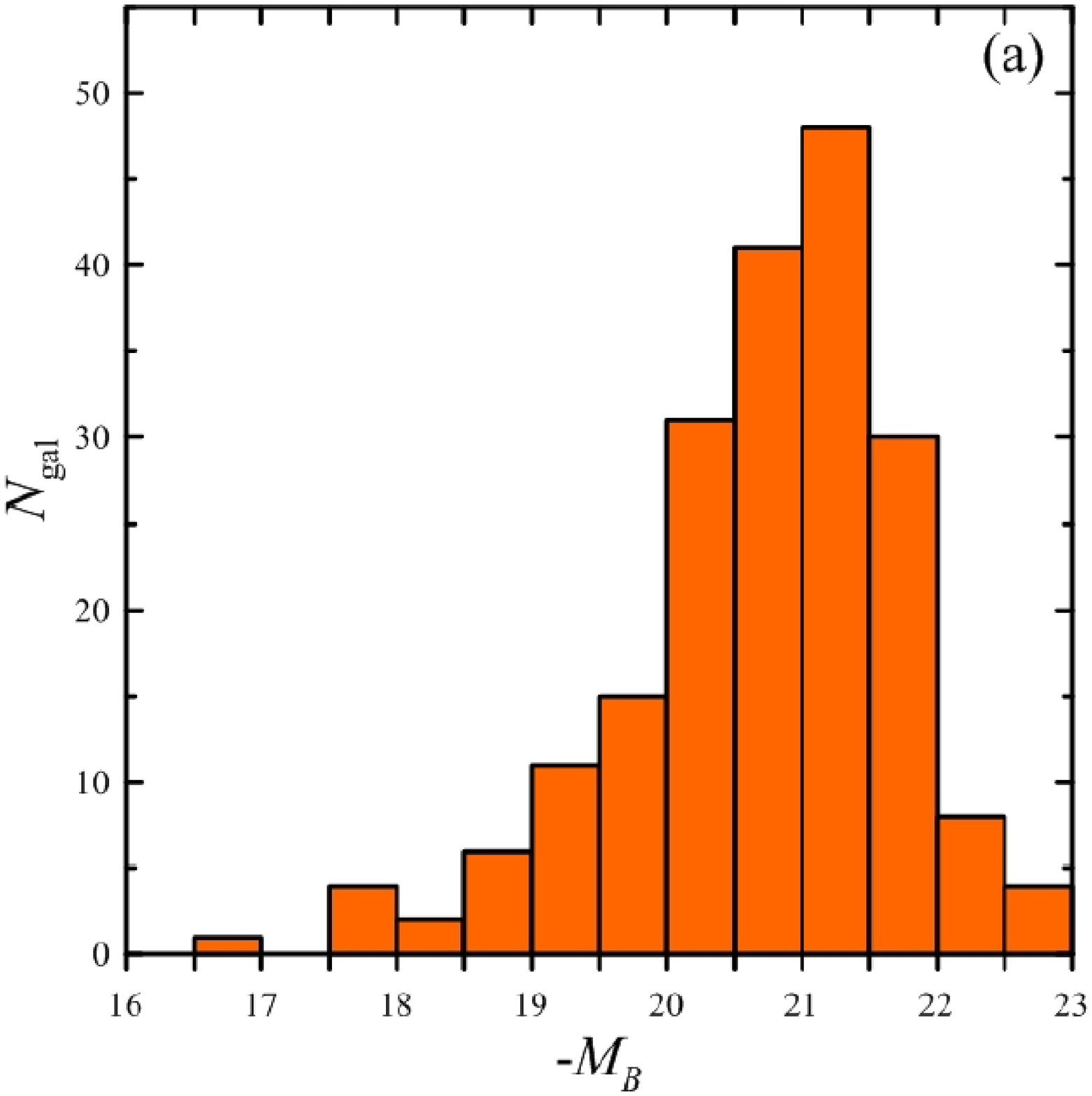} \\ } %
		\end{minipage}
		\hfill
		\begin{minipage}[h]{0.49\linewidth}
			\center{\includegraphics[width=1\linewidth]{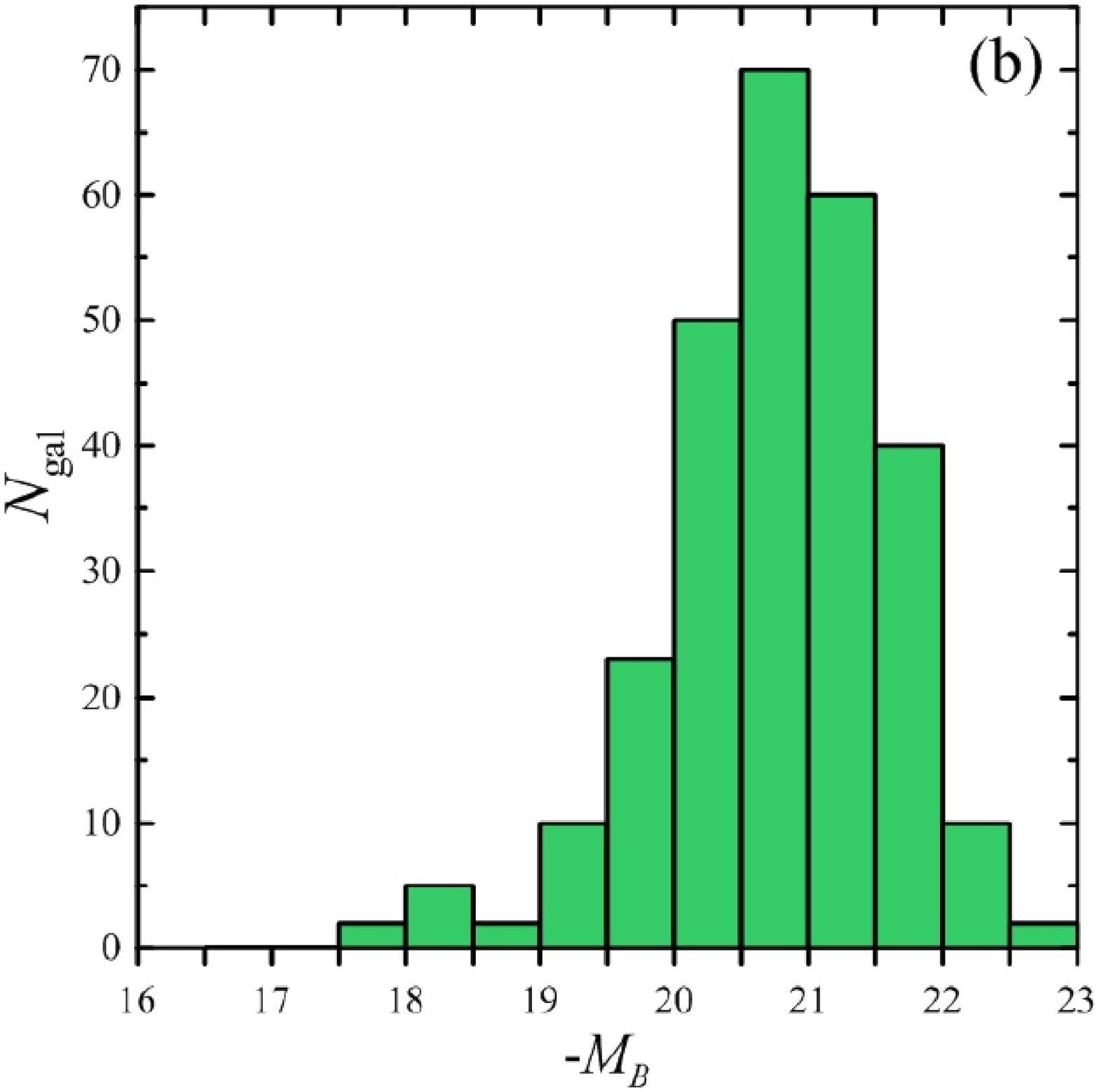} \\ }
		\end{minipage}
		\caption{Distribution of absolute magnitudes of galaxies: (a) for all galaxies of the sample of Chernin et al.~\cite{Chernin_etal_2001}; (b) for our sample
of galaxies.} %% подпись к рисунку
		\label{fig3:MB}
\end{figure*}

Figs.~\ref{fig3:MB}a and~\ref{fig3:MB}b show the
distributions of absolute magnitudes of galaxies for
the two catalogs. The absolute magnitudes of objects
of our sample lie in the interval from $-17^m$ to $-22.5^m$ with a mean of $M_B = 20.5^m$, both for the entire sample
and for barred galaxies, which is consistent with the
results of~\cite{Chernin_etal_2001}.

% Fig 3

\begin{figure*}
	\setcaptionmargin{5mm} \onelinecaptionstrue \captionstyle{normal}
		\begin{minipage}[h]{0.49\linewidth}
			\center{\includegraphics[width=1\linewidth] {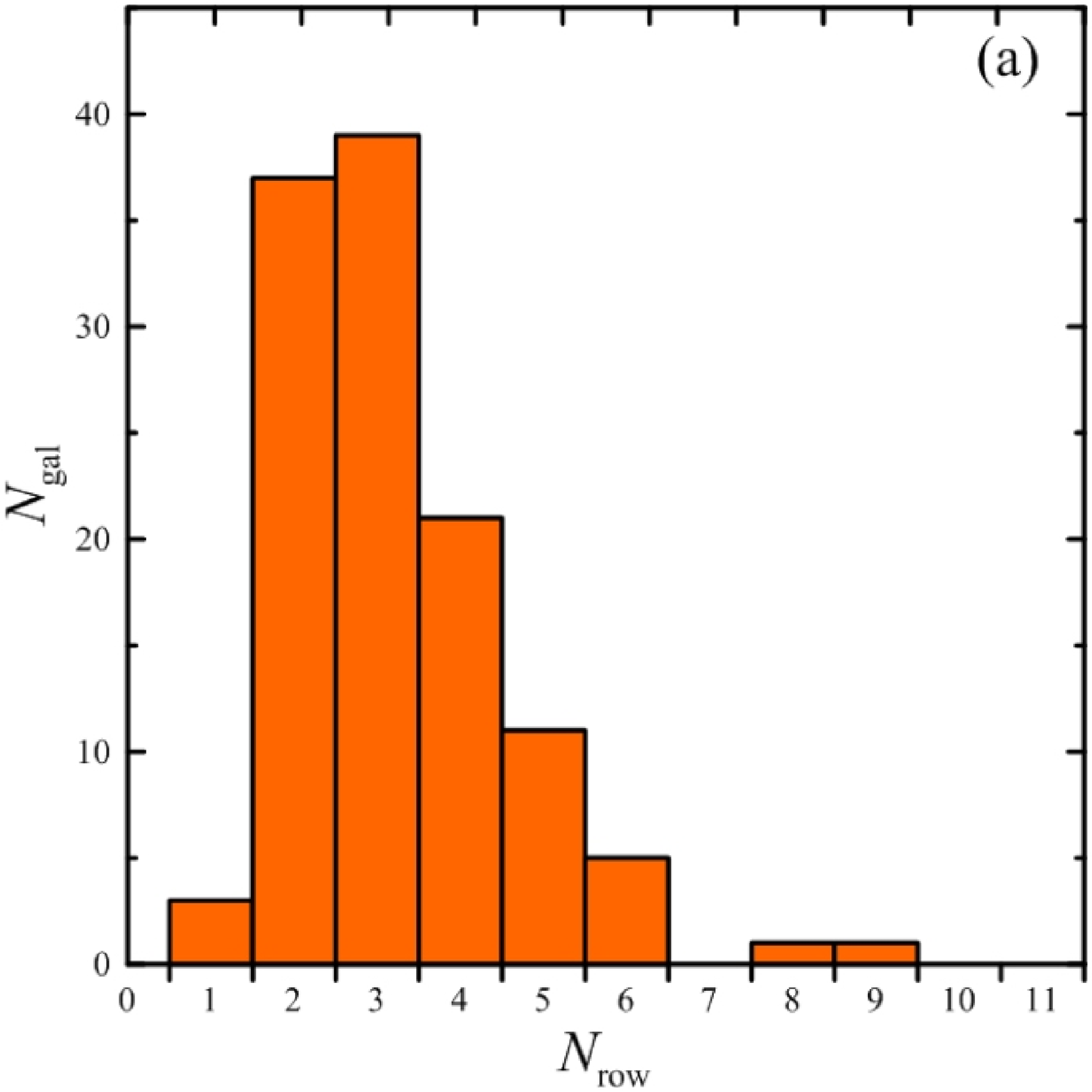} \\} %{N(rows)_new_Chernin.jpg} \\ а)}
		\end{minipage}
		\hfill
		\begin{minipage}[h]{0.49\linewidth}
			\center{\includegraphics[width=1\linewidth]{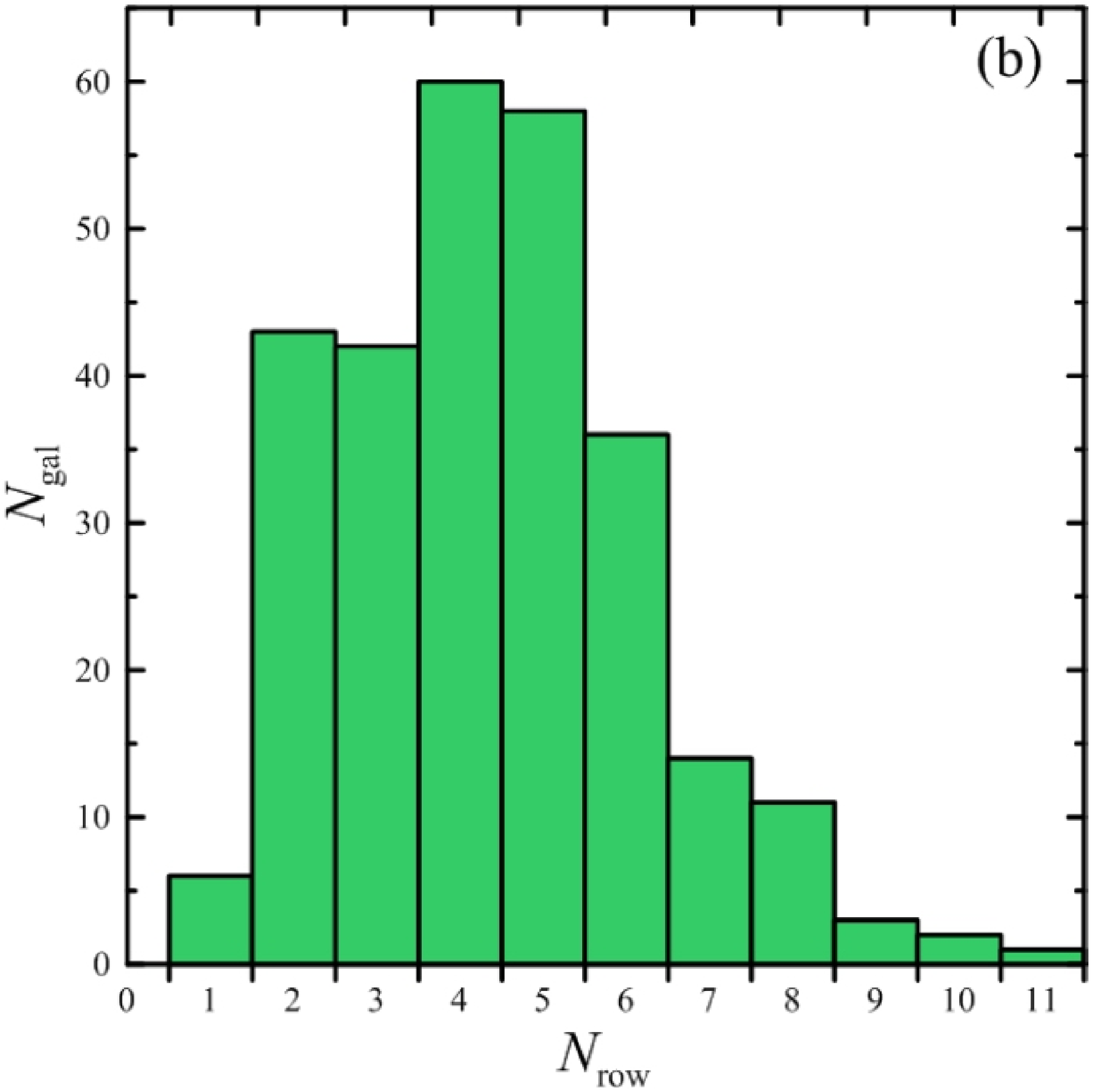} \\}
		\end{minipage}
		\caption{Same as Fig. \ref{fig3:MB}, but for the number of УrowsФ.} %% подпись к рисунку
		\label{fig4:NRows}
\end{figure*}

According HyperLeda data, 77\% of the galaxies of
our sample are barred. The fraction of barred galaxies
in catalog~\cite{Chernin_etal_2001} is equal to 70\% resulting in a 74\% fraction of SB galaxies among the combined sample
of all 406 NGC/IC objects with УrowsФ. However,
HyperLeda indicates the presence of a bar in 51\% of all 7 143 NGC/IC galaxies. The presence of a bar
appears to be a favorable factor for the formation of
polygonal structures in the spiral density wave due
to more suitable conditions for the development of strong galactic shocks (GS) in the stronger potential
of the stellar density wave.
% ≈сли проанализировать по данным LEDA все спиральные галактики, вход€щие в каталог NGC и IC (7143 объектов), то бар присутствует только в 51\% этих объектов. Ќесложно заметить, что частота встречаемости галактик с баром среди галактик с наблюдаемыми полигональными структурами выше, чем среди выборки спиральных галактик из каталога NGC. ѕри рассмотрении большей выборки спиральных галактик процентное отношение галактик с баром уменьшаетс€. »з приведенных результатов анализа можно сделать предположение о том, что наличие бара €вл€етс€ благопри€тным условием дл€ образовани€ таких транзиентных структур в спиральной волне плотности.
The number of УrowsФ $N$ in galaxies of our catalog
varies from one to 11 with a mean of $\langle N\rangle = 4$ (Fig. \ref{fig4:NRows}). In the case of catalog~\cite{Chernin_etal_2001} the average number
of УrowsФ for the entire sample is close to three with
two or three УrowsФ found inmost of the galaxies. The
distribution of the number of УrowsФ in our sample is
rather broad with a less pronounced maximum (see
Fig. \ref{fig4:NRows}б). Like in the catalog of Chernin et al.~\cite{Chernin_etal_2001}, we
found no correlation between the number of УrowsФ
and absolute magnitude $M_B$.

% Fig 4

\begin{figure*}
	\setcaptionmargin{5mm} \onelinecaptionstrue \captionstyle{normal}
	\begin{minipage}[h]{0.49\linewidth}
		\center{\includegraphics[width=1\linewidth] {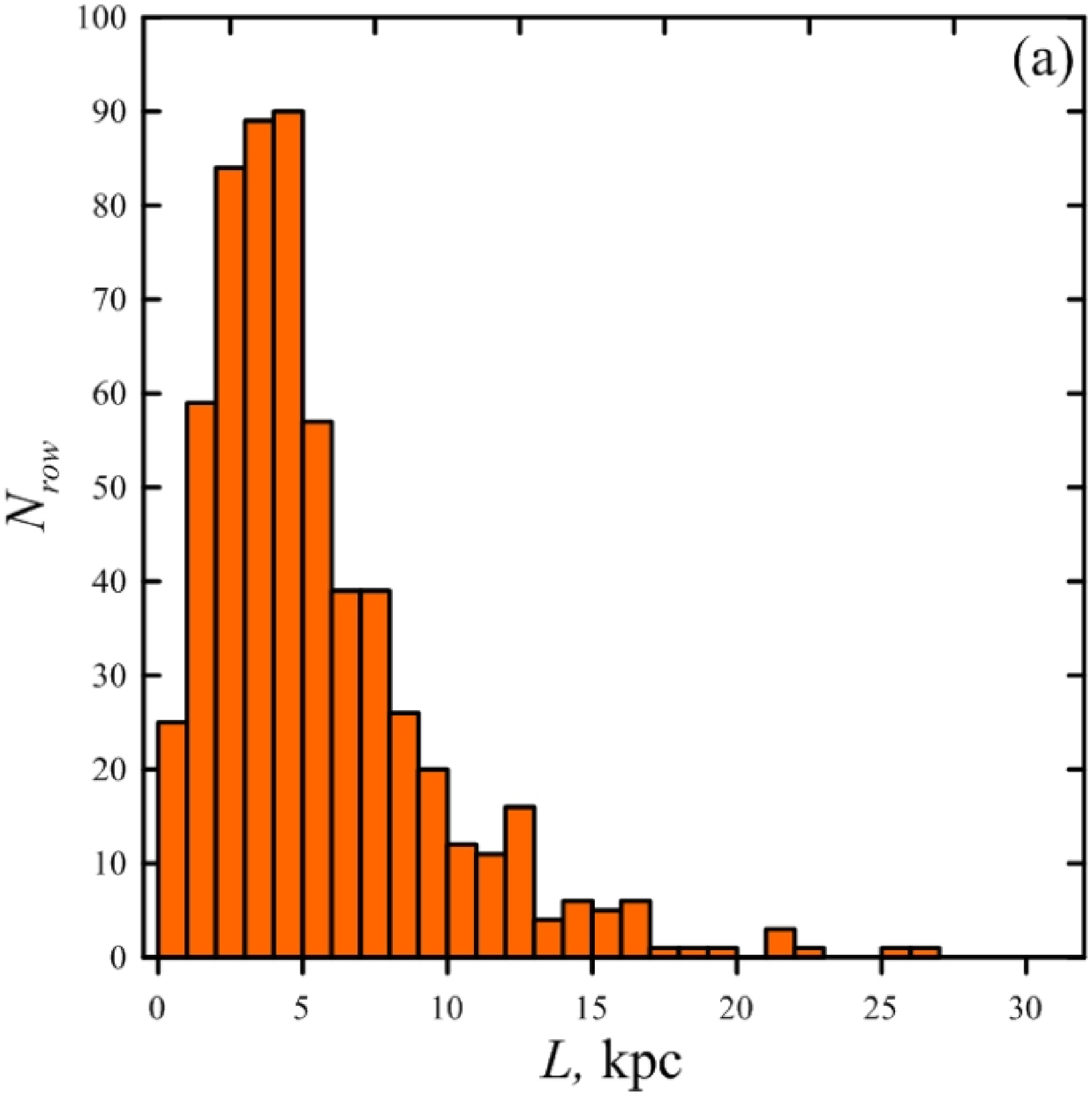} \\}
	\end{minipage}
	\hfill
	\begin{minipage}[h]{0.49\linewidth}
		\center{\includegraphics[width=1\linewidth]{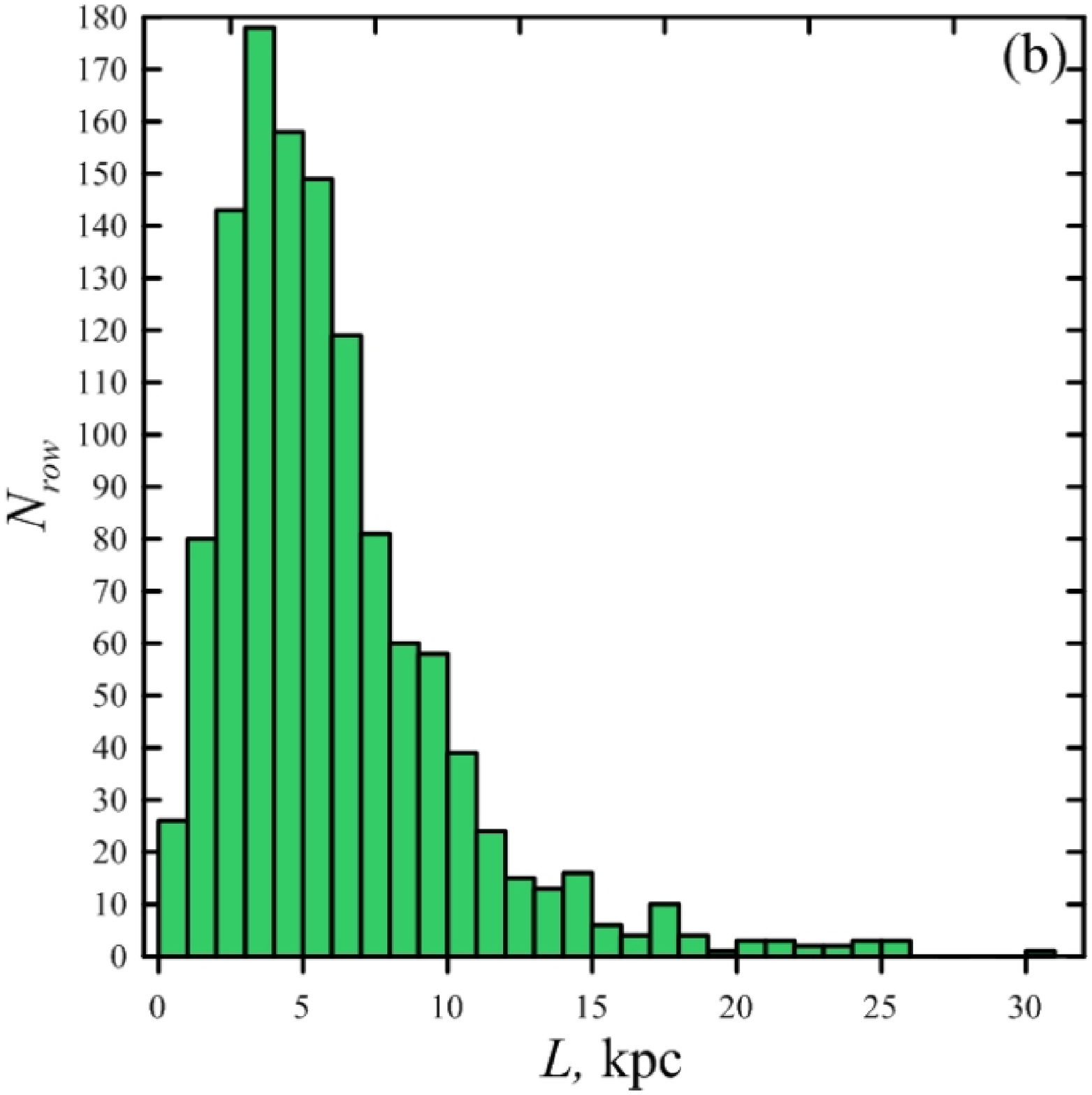} \\}
	\end{minipage}
	\caption{Distribution of linear sizes of УrowsФ: (a) for all galaxies from the sample of Chernin et al. \cite{Chernin_etal_2001}; (b) for our sample of
galaxies.} %% подпись к рисунку
	\label{fig5:LRows}
\end{figure*}

Linear sizes of УrowsФ, $L$, vary over a wide range,
exceeding 22 kpc in several galaxies, e.g., NGC~0010, NGC~1365, NGC~1512, NGC~3976, NGC~5850, NGC~6935, and reaching 30 kpc in IC~4479 (Fig.~\ref{fig5:LRows}). The average length of a УrowФ is $\langle L\rangle=6.6$~kpc, and
the median length is 5.6 kpc. These results are almost
twice the values for catalog~\cite{Chernin_etal_2001}. This discrepancy is
due to (1) higher fraction of long УrowsФ because
our analysis reveals more extended spiral structure
and (2) because of the increase of $L$ with galactocentric
distance $r$ combined with high quality of
SDSS images of peripheral parts of galaxies and the
fact GALEX images allow tracing spiral structure far
beyond the optical radius.

% Fig 5

\begin{figure*}[]
	\setcaptionmargin{5mm} \onelinecaptionstrue \captionstyle{normal}
	\includegraphics[scale=0.6]{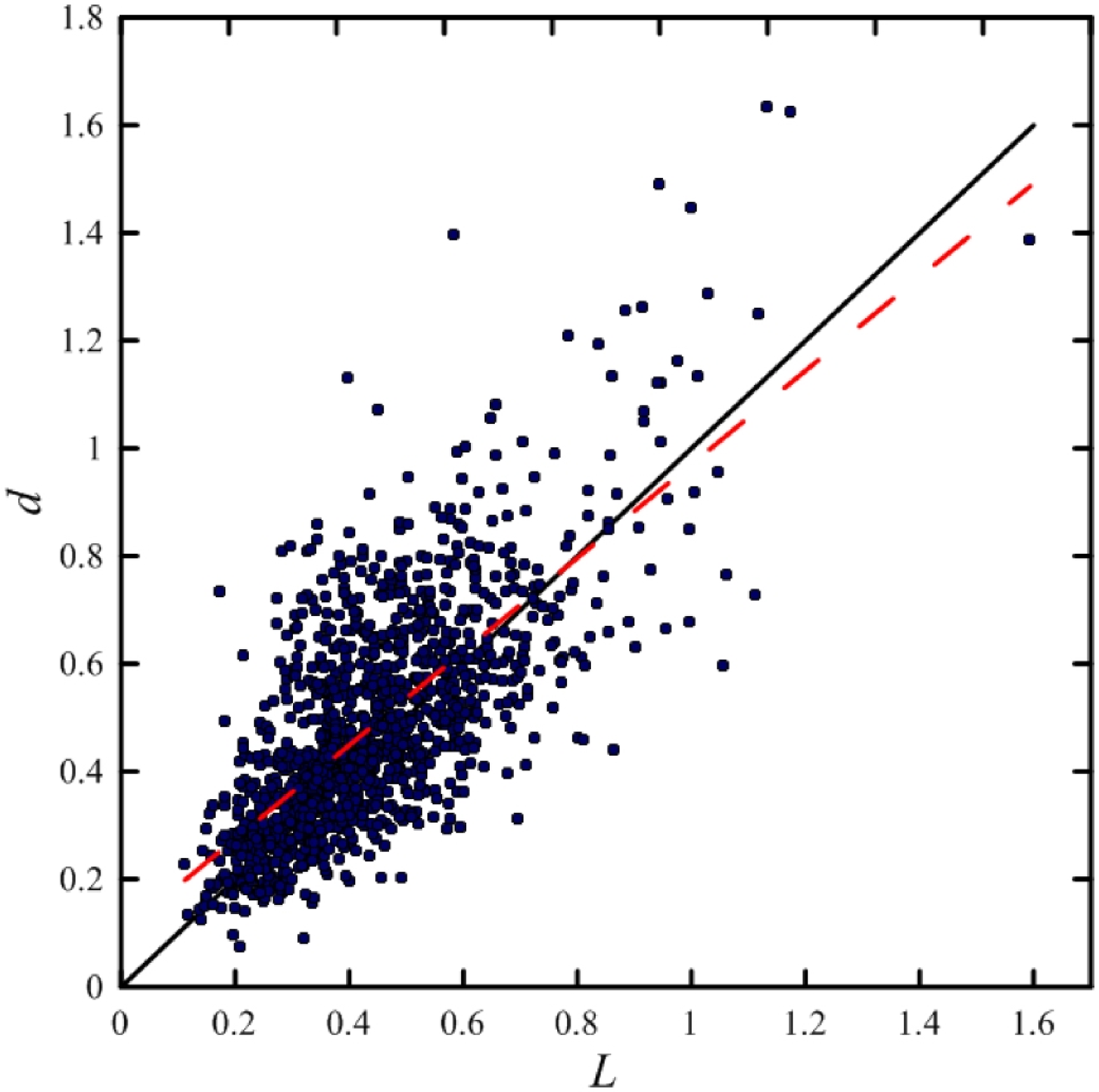}
	\caption{The Уrow length $L$ --- galactocentric distance $d$ diagram (both quantities are normalized to the optical
radius of the galaxy, $D_{25}/2$). The dashed and solid lines
show the regression relation and the identity relation $d = L$, respectively}
	\label{fig6:L_d}
\end{figure*}

Like in the case of catalog~\cite{Chernin_etal_2001}, we find the length of
the УrowФ to correlate with galactocentric distance $d$ of the end of the УrowФ: the farther from the center,
the longer the УrowsФ (Fig.~\ref{fig6:L_d}). The length of the
УrowФ and galactocentric distance $d$ are normalized
to the optical radius of the galaxy, $D_{25}/2$. The solid
line in Fig.~\ref{fig6:L_d} shows the dependence $d = L$, and the
dashed line, the linear approximation of this dependence, ($d = 0.87 L + 0.1$).
%ќтклонение от закона в нашем случае $d = (1 \pm  0.31) L$  сопоставимо с оценкой $d = (1 \pm  0.11) L$ дл€ каталога~\cite{Chernin_etal_2001}.

%Fig 6

\begin{figure*}
	\setcaptionmargin{5mm} \onelinecaptionstrue \captionstyle{normal}
	\begin{minipage}[h]{0.49\linewidth}
		\center{\includegraphics[width=1\linewidth]{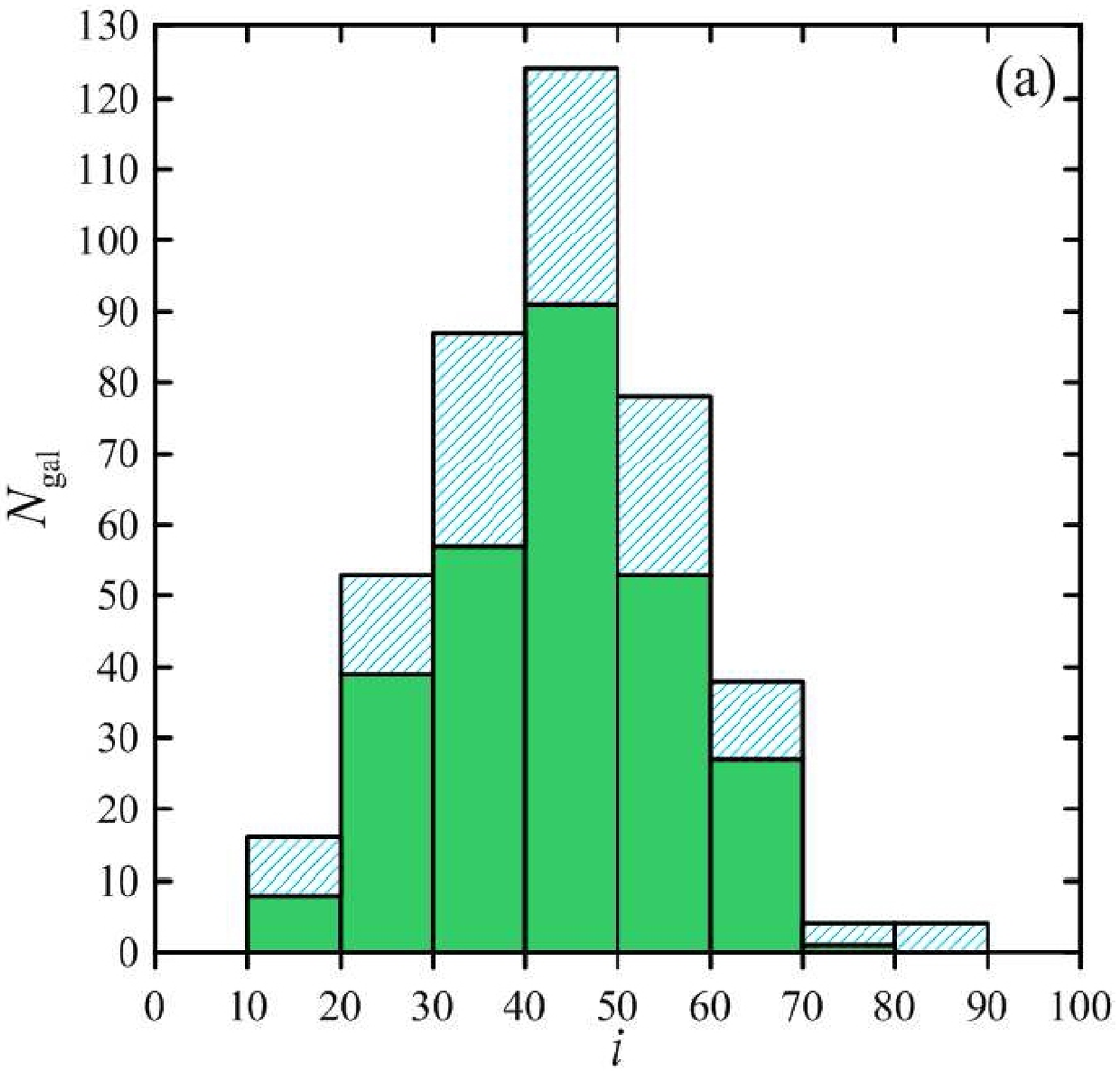} \\}
	\end{minipage}
	\hfill
	\begin{minipage}[h]{0.49\linewidth}
		\center{\includegraphics[width=1\linewidth]{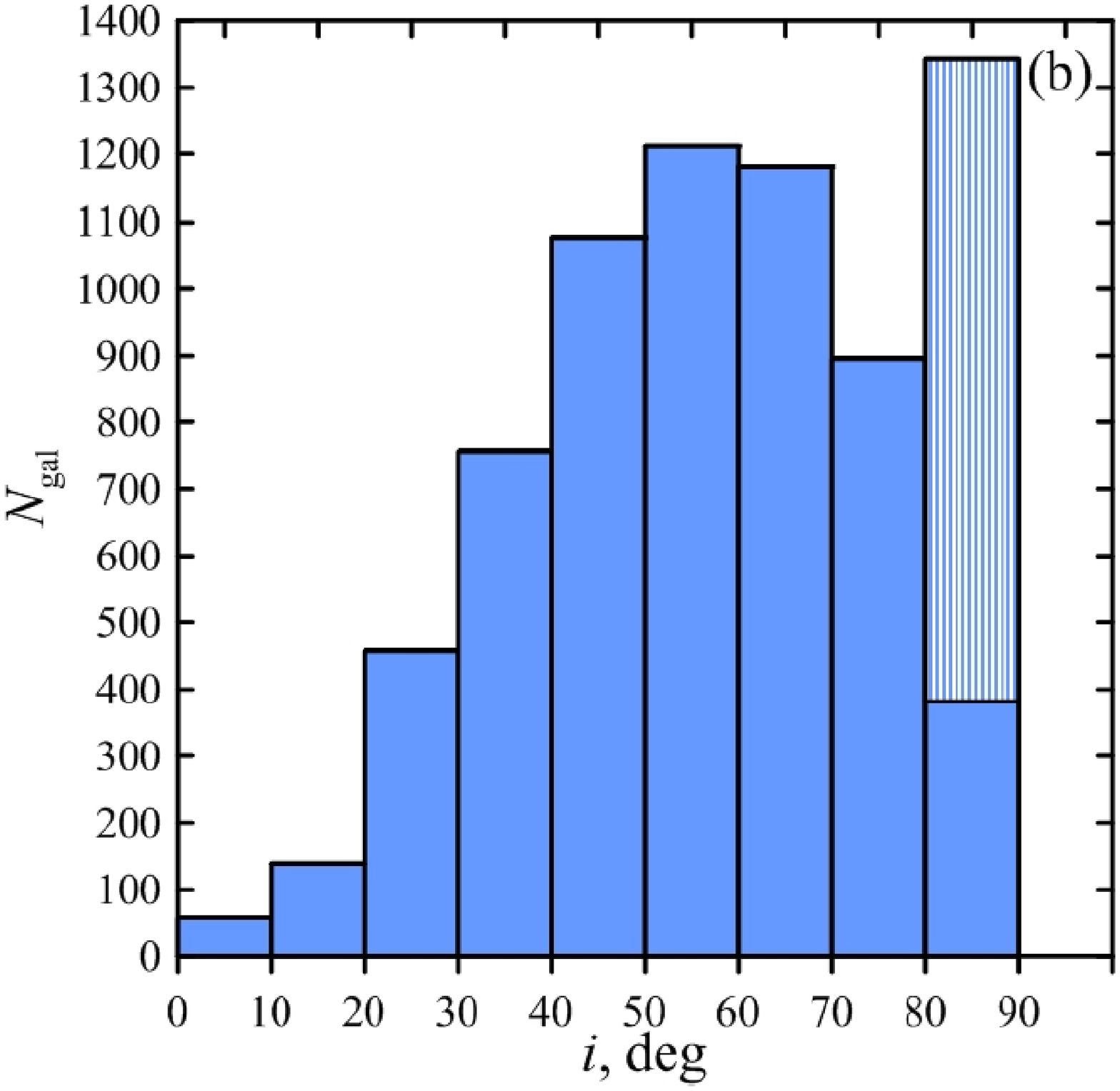} \\}
	\end{minipage}
	\caption{Distribution of inclinations $i$ of spiral galaxies: (a) for NGC and IC galaxies with УrowsФ (276 objects of our catalog --
the shaded bars, 406 objects including NGC objects from the catalog of Chernin et al. \cite{Chernin_etal_2001} -- заштрихованные столбцы); the dashed bars); (b) for all spiral
galaxies of the NGC/IC catalogs (according to HyperLeda data); the stripped barЧfir those with $i\geq 90^\circ$}
	\label{fig12:N(i)}
\end{figure*}

Figure \ref{fig12:N(i)} shows the distributions of inclination i for
galaxies with УrowsФ (7a) and for all S-type NGC and
IC galaxies according to HyperLeda data. The entire
sample of NGC/IC galaxies contains a large fraction
of objects with inclinations $i\simeq 90^\circ$, which is due to
the effect of higher surface brightness of edge-on
galaxies resulting from geometric factor. There is also
an extra contribution due to $i=90^\circ$ being assigned to
some non-disk galaxies in the HyperLeda database.
The fraction of polygonal galaxies with inclinations $i > 70^\circ$ is small because of the problems with the
determination of the properties of the spiral pattern,
and we exclude these objects from consideration.

% Fig 7

Figure~\ref{fig12:N(i)} demonstrates the qualitative similarity of
galaxies with УrowsФ and all S-type NGC and IC galaxies as far as the distribution of inclinations $i$is concerned. In both cases we have a bell-shaped
distribution with a maximum near $40^\circ$ for galaxies
with УrowsФ and $50^\circ$for all S-type galaxies. The nonuniformity
of the distribution is due to observational
selection and the method used to determine inclination
based on the photometric axis ratio $b$:$a$. Because
of the presence of the bar, bulge, and various shape
distortions in interacting galaxies and objects of late
morphological types such axial-ratio based estimates
should evidently result in systematic deviations from
uniform distributions both for galaxies with small
(face-on) and large (edge-on) inclinations $i$ \cite{Karachentseva_2016}.

\begin{figure*}
	\setcaptionmargin{5mm} \onelinecaptionstrue \captionstyle{normal}
	\begin{minipage}[h]{0.49\linewidth}
		\center{\includegraphics[width=1\linewidth]{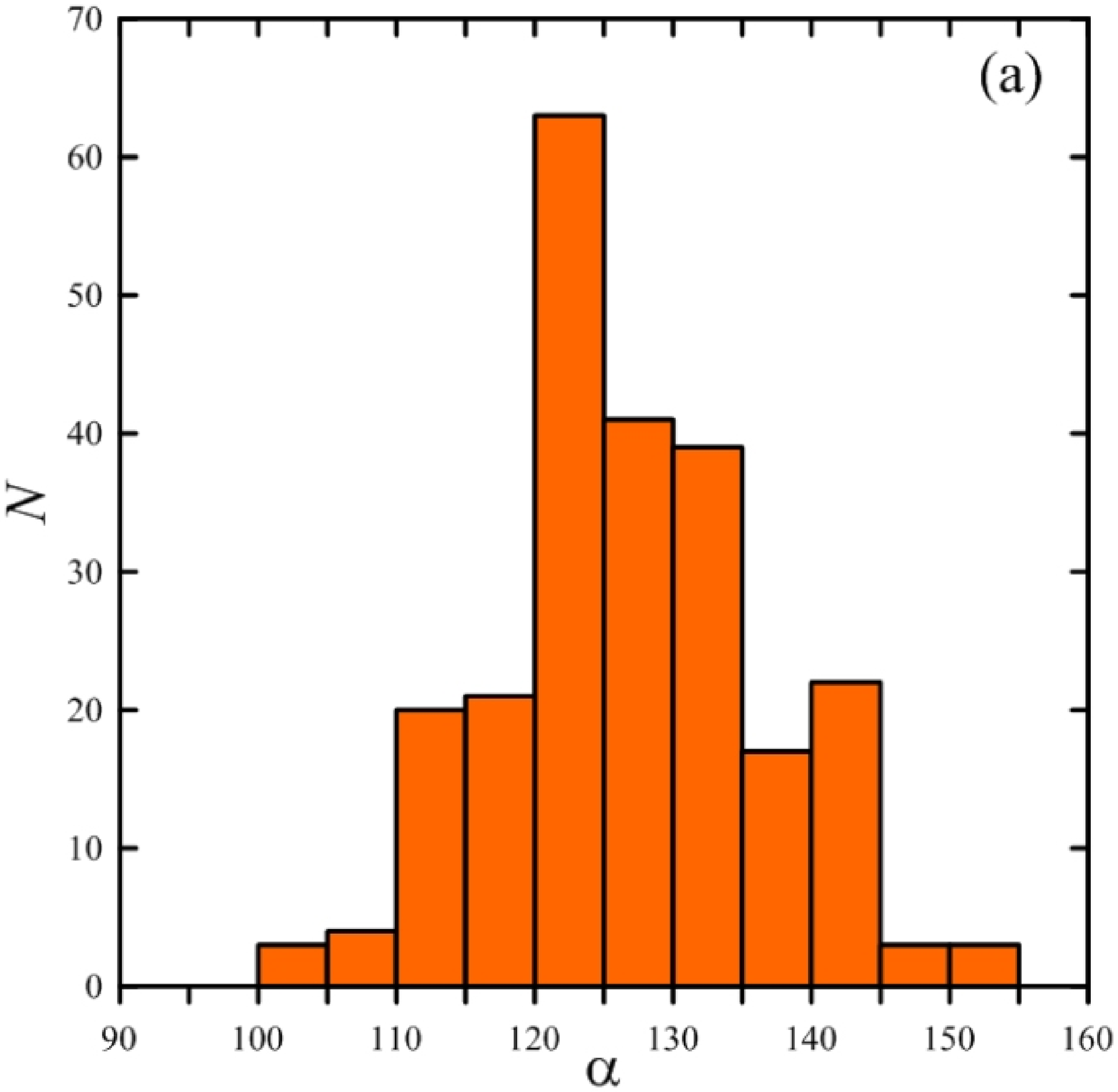} \\}
	\end{minipage}
	\hfill
	\begin{minipage}[h]{0.49\linewidth}
		\center{\includegraphics[width=1\linewidth]{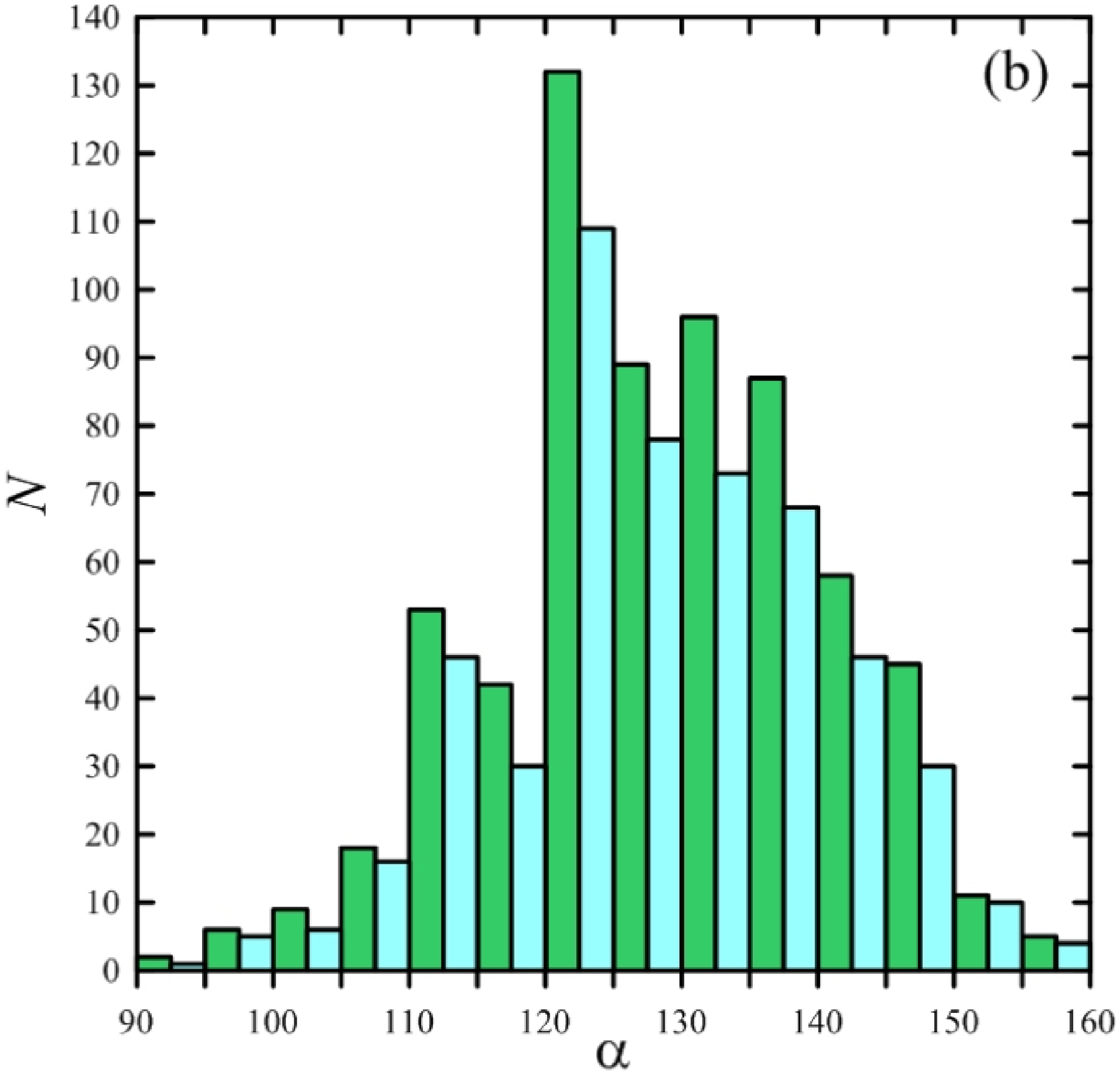} \\}
	\end{minipage}
	\caption{Histogram of the distribution of angles $\alpha$ between the adjacent УrowsФ: (a) for all galaxies of sample \cite{Chernin_etal_2001}; (b) for our
sample of galaxies. The light-shaded bars show the distribution for galaxies with $i < 55^\circ$} %% подпись к рисунку
	\label{fig7:Angle}
\end{figure*}

Straight-line segments are difficult to identify in
galaxies with large inclination and that us why, given the incompleteness of our sample, hereafter for inclinations $i\geqslant 55^\circ$ (see Fig.~\ref{fig12:N(i)}), we show the distributions
for both the combined sample of galaxies with УrowsФ,
and separately for galaxies with $i< 55^\circ$.
%далее будем нар€ду с полной выборкой рассматривать отдельно галактики с углом наклона $i < 55^\circ$.
%»з рисунка~\ref{fig12:N(i)} также видно, что выборка галактик с вереницами неполна начина€ с $i\geqslant 55^\circ$, поскольку в сильно наклоненных галактиках сложно выделить спр€мленные участки в спиральном узоре галактики.
%”читыва€ эту неполноту нашей выборки, далее будут приведены распределени€ как по полной выборке галактик с вереницами так и отдельно дл€ галактик с $i< 55^\circ$.
The angle between the adjacent УrowsФ is usually
close to $\alpha=120^{\circ}$  (Fig.~\ref{fig7:Angle}). The average angle is
equal to $126^{\circ}$, both for our entire sample and for
barred galaxies, in agreement with the result of
Chernin et al.~\cite{Chernin_etal_2001}.
%, однако медианное значение угла $\alpha$ несколько выше, чем в~\cite{Chernin_etal_2001}.

% Fig 8

\begin{figure*}
	\setcaptionmargin{5mm} \onelinecaptionstrue \captionstyle{normal}
	\begin{minipage}[h]{0.325\linewidth}
		\center{\includegraphics[width=1\linewidth]{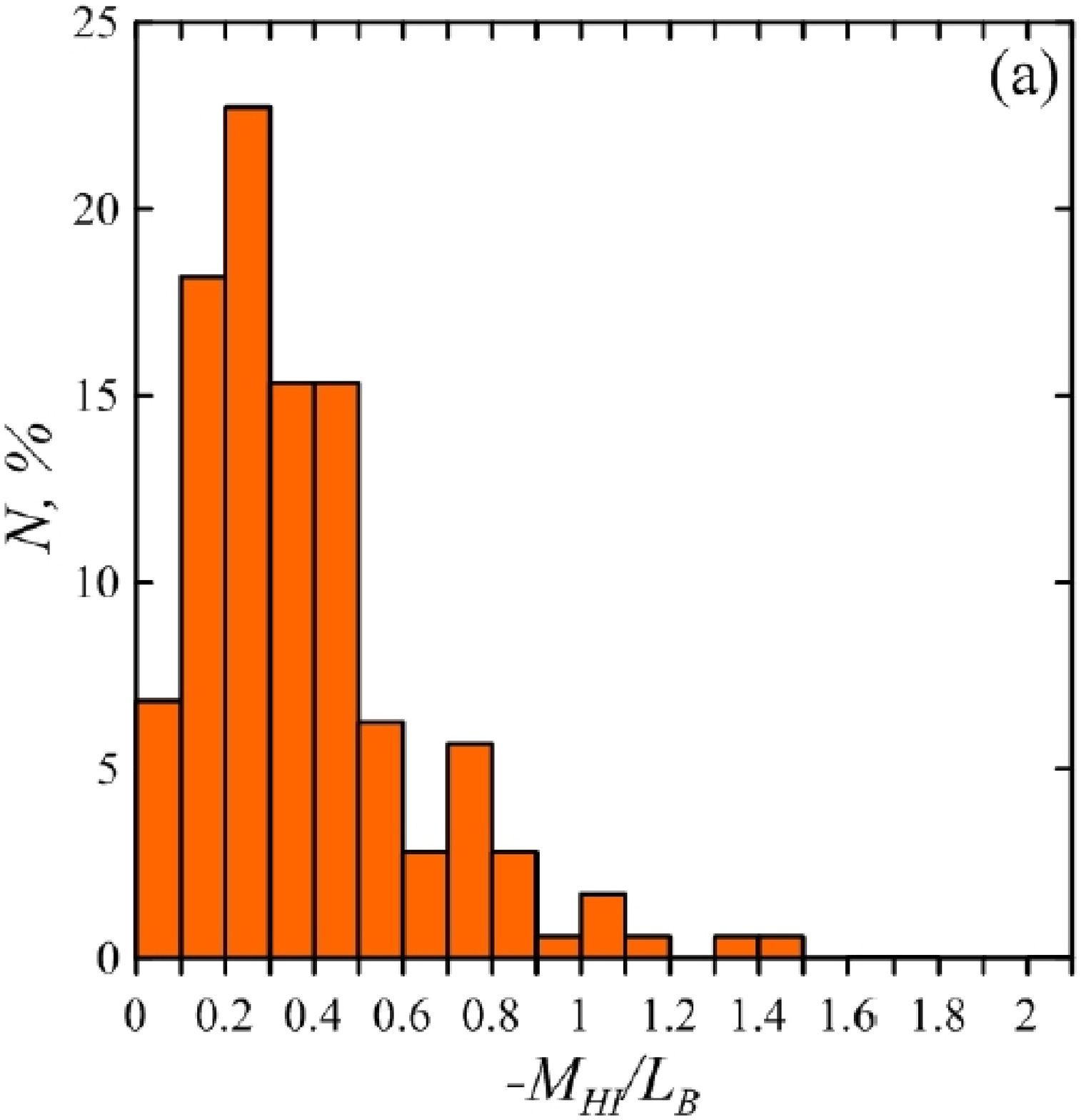} \\} %{N_Mtipe_new_Chernin.jpg} \\ а)}
	\end{minipage}
	\hfill
	\begin{minipage}[h]{0.325\linewidth}
		\center{\includegraphics[width=1\linewidth]{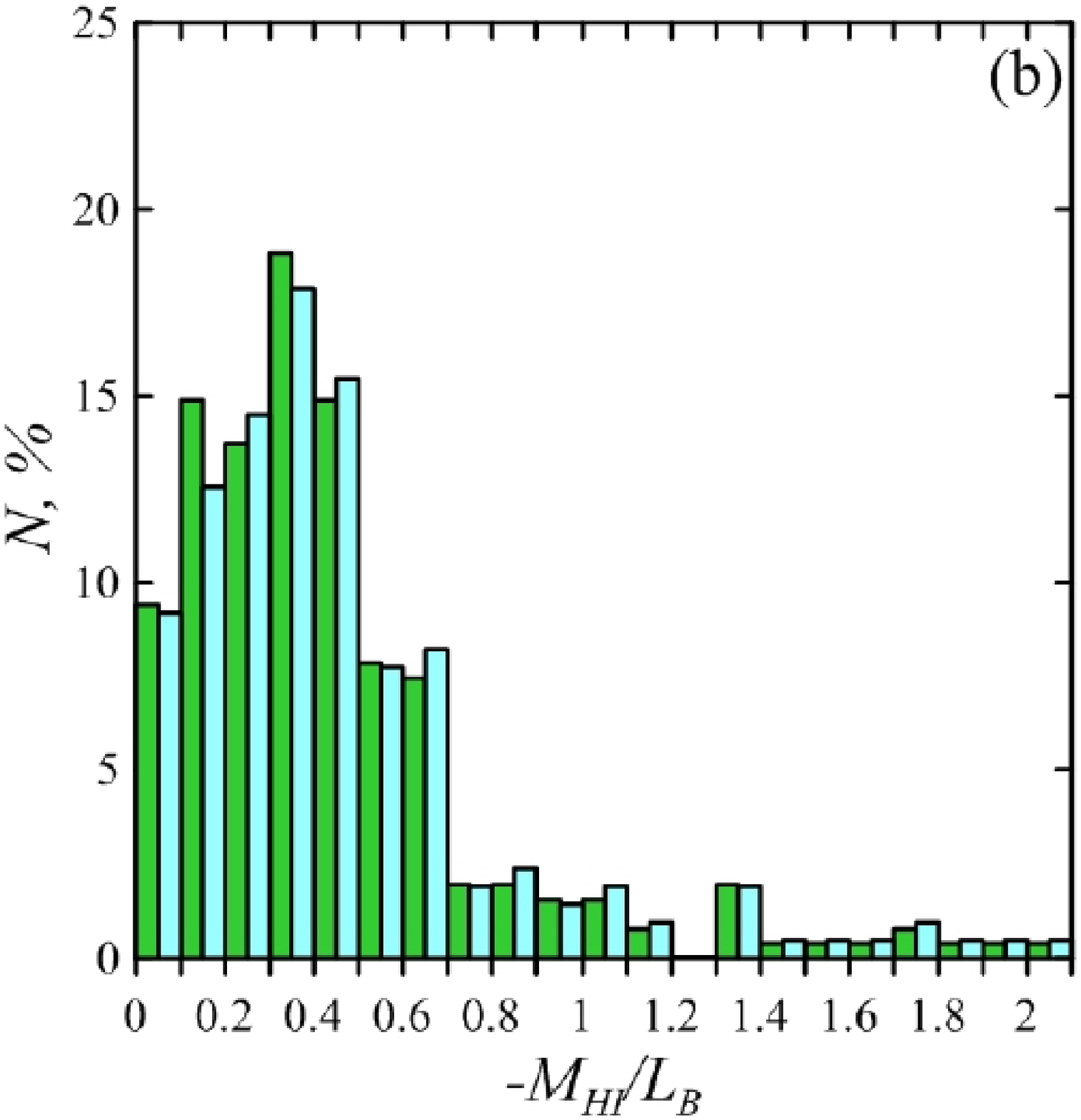} \\}
	\end{minipage}
	\hfill
	\begin{minipage}[h]{0.325\linewidth}
		\center{\includegraphics[width=1\linewidth]{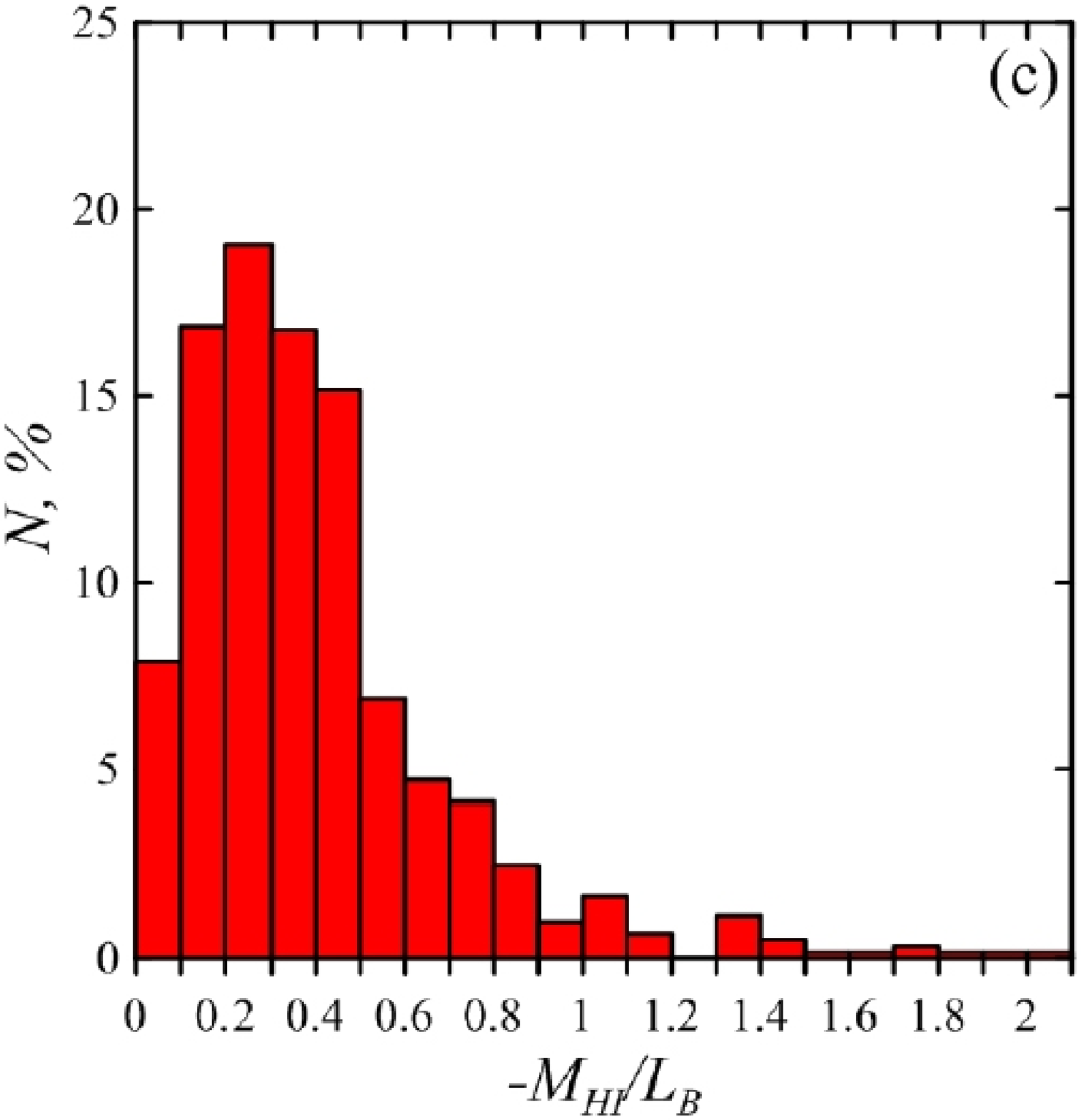} \\}
	\end{minipage}
\caption{Distribution of the $-M_{HI}/L_B$ ratio (in solar units) for galaxies with УrowsФ: (a) for the full sample of Chernin et al. \cite{Chernin_etal_2001} updatedwith data from HyperLeda; (b) for our sample (the light-shaded bars show the distribution for galaxieswith inclinations $i< 55^\circ$; (c) the combined histogram for all objects with polygonal structures.}
\label{fig8:MHI}
\end{figure*}

Figure~\ref{fig8:MHI}shows the estimates of gas content in
galaxies with УrowsФ: the distributions of the $M(HI)/L_B$ ratio (in solar units). The maximum is at 0.3 and the
mean ratio for the entire sample is $\langle M(HI)/L_B\rangle=0.39$ which is consistent with the results of~\cite{Chernin_etal_2001}.

% Fig 9

\begin{figure*}[]
	\setcaptionmargin{5mm} \onelinecaptionstrue \captionstyle{normal}
	\includegraphics[scale=0.6]{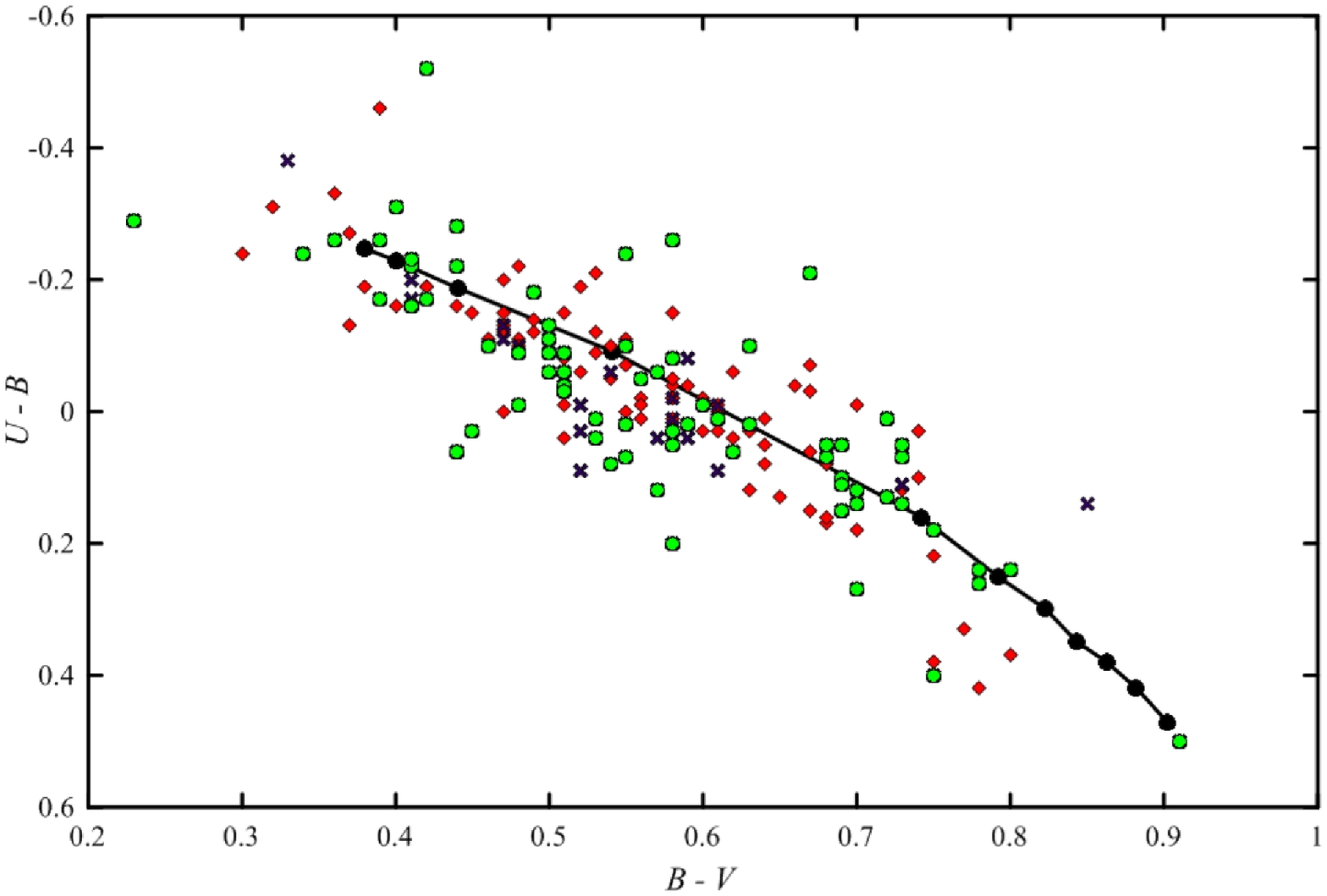}
	\caption{Color-color diagram for galaxies with УrowsФ:
the diamond symbols show the data for the full sample of
Chernin et al. \cite{Chernin_etal_2001}; the circles and crosses show the objects
of our sample with $i<55^\circ$, and $i\geqslant 55^\circ$, respectively. The
color indices are corrected for selective Galactic extinction
and galaxy disk inclination. Actual data from HyperLeda are used.}
	\label{fig11:UBV}
\end{figure*}

According to HyperLeda, UBV photometry with
color indices corrected for selective Galactic extinction
and galaxy disk inclination is available for 34\% of
the galaxies with УrowsФ from our catalog. Figure~\ref{fig11:UBV} shows the $(U - B)_0Ц(B - V)_0$, color-color diagram, where diamond symbols indicate updated data or the
sample of Chernin et al.~\cite{Chernin_etal_2001}, and the crosses and
circles show the positions of objects of our sample.
The data points for all galaxies with УrowsФ agree well
with the intrinsic color relation, which is shown by the
solid line.
%$(U - B)_0Ц(B - V)_0$

%Fig 10

\section{FORMATION MECHANISMS OF
POLYGONAL STRUCTURES}

Two possible formation mechanisms for polygonal
structures have been discussed in the literature.
Chernin \cite{Chernin_1999A&AT,Chernin_1999MNRAS} proposed hydrodynamical mechanism
based on particularities of the dynamics of the
global galactic shock. Because of the instability of
the shock front it tends to become locally flat and
polygonal structure forms. This mechanism is bona
fide reproduced in numerical simulations of various
authors \cite{Khoperskov_2011_polygon_str,Khoperskov_2012_polygon_str,Filistov_2012,Filistov_2015,Butenko_Khoperskov_S_2011}.

Another mechanism is also possible where individual
straight-line arm segments can be recognized
in numerical N-body models at the stage of
the formation of the transient bar \cite{Buta_Combes_1996}. Models based
on the 4/1 resonance can also describe the observed
features of the spiral structure \cite{Contopoulos_Grosbol_1986, Patsis_Grosbol_1997}. The mechanism
of the formation of УrowsФ based on the gravitational
instability of the collisionless stellar disk and/or on resonance phenomena was analyzed by
Rautiainen et al. \cite{Rautiainen_etal_2008,Rautiainen_Melnik_2010,Melnik_Rautiainen_2013}. When constructing
their numerical models of stellar disks the above authors
used gravitational potentials constructed from
H-band photometry of the OSUBSGS (Ohio State
University Bright Spiral Galaxy Survey) survey. They
could reproduce straight-line segments in terms of
a collisionless N-body model, in particular, for the
NGC~4303 galaxy.
%—ами авторы \cite{???} указывают на роль резонансных эффектов.

\begin{figure*}
	\setcaptionmargin{5mm} \onelinecaptionstrue \captionstyle{normal}
	\begin{minipage}[h]{0.49\linewidth}
		\center{\includegraphics[width=1\linewidth]{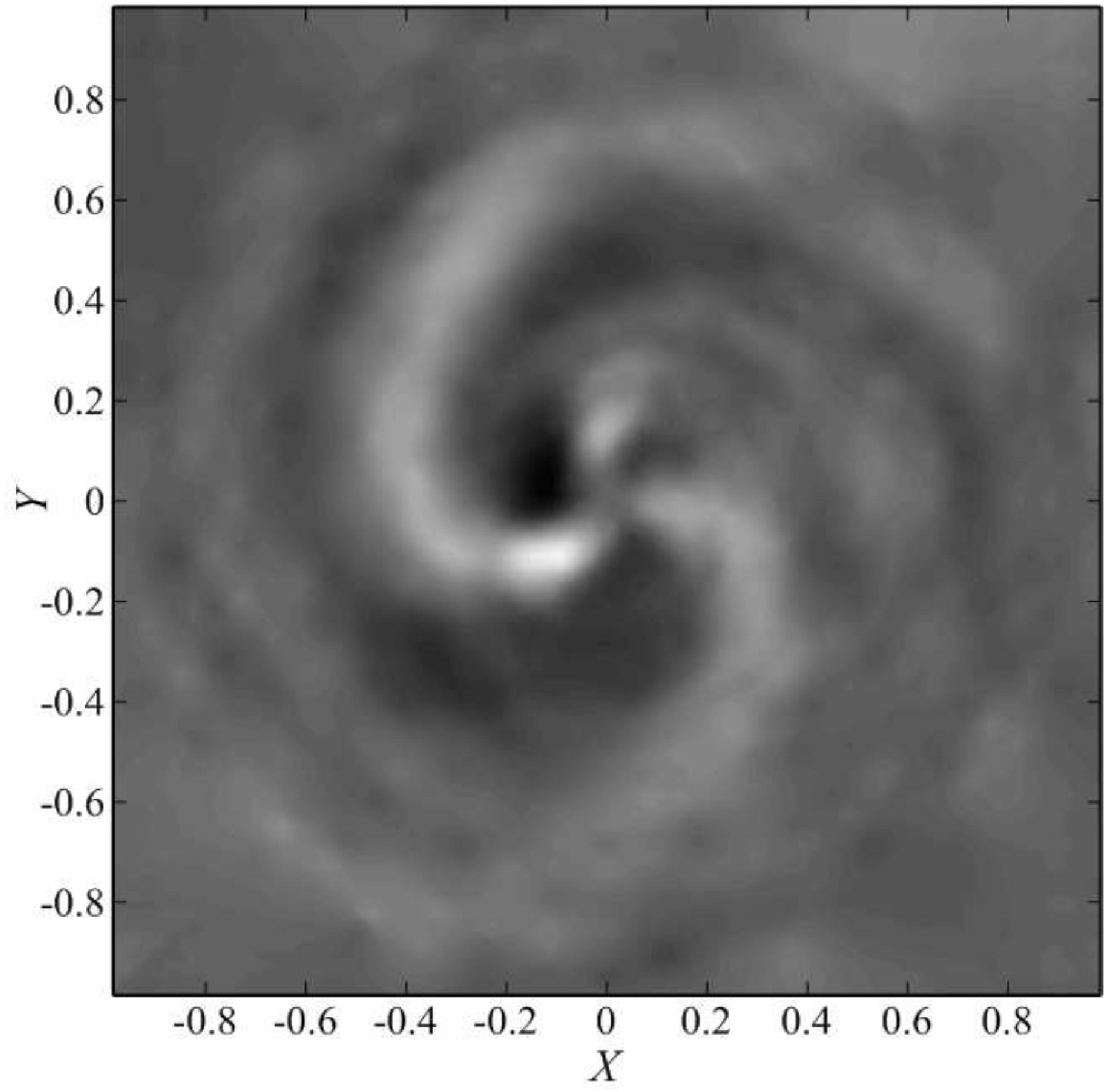} \\ }
	\end{minipage}
	\hfill
	\begin{minipage}[h]{0.49\linewidth}
		\center{\includegraphics[width=1\linewidth]{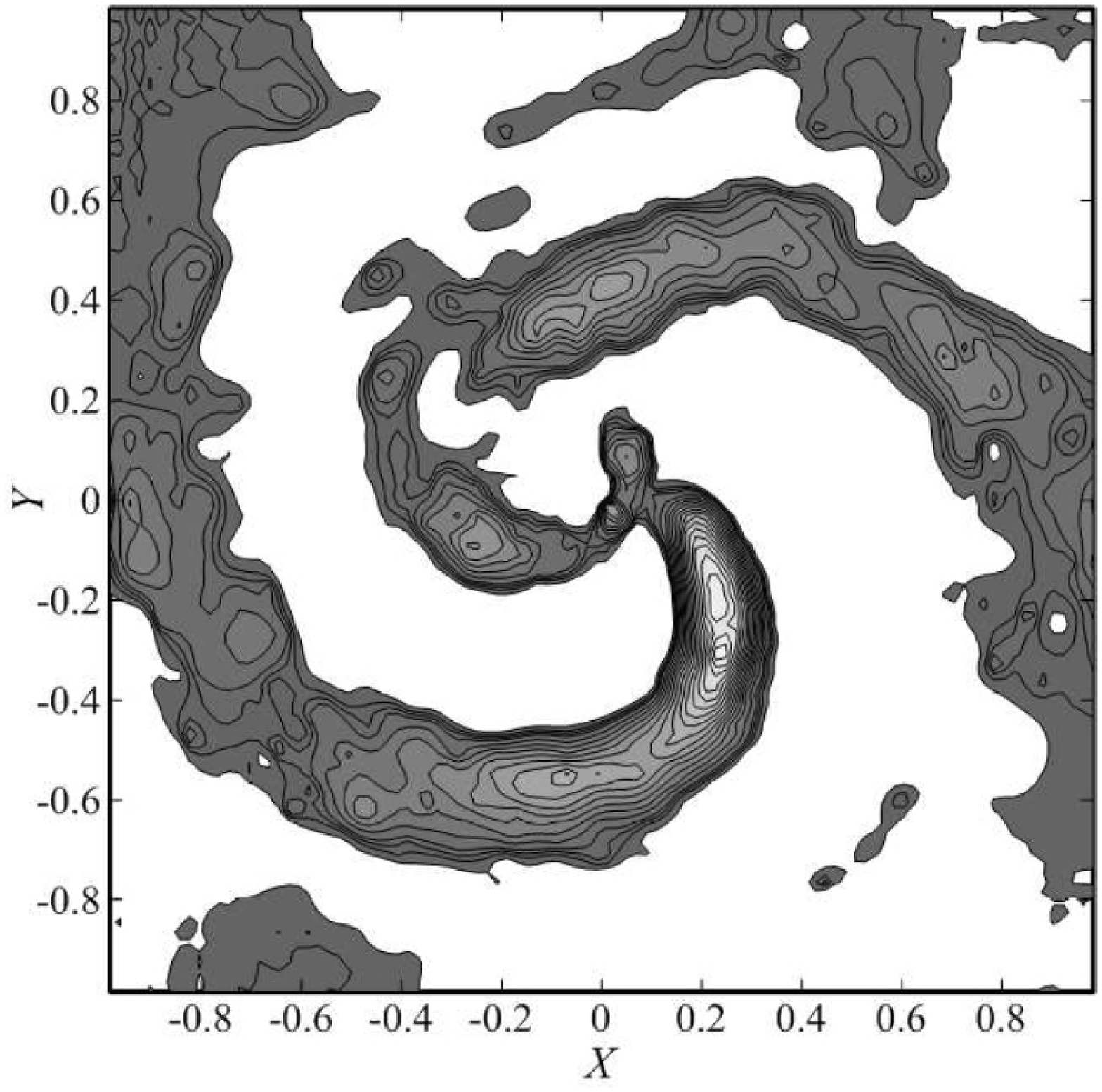} \\ }
	\end{minipage}
	\caption{Distribution of the perturbation of surface density logarithm in the model stellar disk at different time instants (the
left-hand panel shows only the positive part of the perturbation). At the initial time instant the Toomre parameter in the region $1/2\lesssim r/r_d \lesssim 2$ is $Q_T=0.8$, where $r_d$ -- is the exponential disk scale.} %% подпись к рисунку
	\label{fig13:N-boby_model}
\end{figure*}

Let us now point out some distinctive features
in stellar disks according to the results of our dynamic
N-body simulations without going into details
of particular physical mechanisms of the formation of
УrowsФ \cite{Rautiainen_Melnik_2010,Melnik_Rautiainen_2013}. A detailed description of the numericalmodel
employed can be found in \cite{Khoperskov_Bizyaev_2010, KhoperskovSA_KhoperskovAV_2012}. Straightline
segments appear in models with sufficiently massive
halos with $\mu = M_h/M_d \gtrsim 3$ within the optical
radius. Only in such models multi-armed and sufficiently
narrow spirals with recognizable transient
straight-line segments (УrowsФ) can form at various
values of the Toomre parameter $Q_T \lesssim 1$. Figure \ref{fig13:N-boby_model} shows the results of our numerical N-body models
where straight-line arm segments and even almost
kink points can be seen. Let us point out some of the
features of models of stellar disks with УrowsФ:
\begin{itemize}
	\item[--]{Straight-line spiral segments are non-stationary,
like in the case of УrowsФ in a gaseous disk. However, their lifetime scales in hot collisionless
medium are appreciably shorter than in gas;}
	\item[--]{These structures arise rather seldom and for a
short time, usually during the initial stages of the
development of gravitational instability;}
	\item[--]{Polygonal structure consists of one to three
УrowsФ, unlike what we have in gas-dynamic
models, where a global system of УrowsФ forms.}	
\end{itemize}

% Fig 11

% Ќа рисунке \ref{fig13:N-boby_model} приведены результаты моделировани€ бесстолкновительного диска, гравитационно неустойчивого в начальный момент времени.
 %Ќа рисунке \ref{} показано распределение возмущени€ поверхностной плотности.

We already pointed out that according to 2MASS
data, УrowsФ occur rather rarely in old stellar disks
(see Subsection \ref{subsection_Character}).
 Straight-line segments can be
recognized in 2MASS images of 18 objects of our
sample, and this fact is indicative of the large amplitude
of the spiral wave in the massive stellar disk. One would hardly expect perturbation in gas (whose
mass is, on the average, small compared to the mass
of stars) to be able to build up a УrowФ in the old stellar
component. The problem is further complicated by
the non-stationary (transient) nature of the polygonal
structure: under these conditions the time scale of the
gravitational influence of the gaseous structure on the
stellar component may be insufficient. In the case of
the 13 galaxies mentioned above the primary formation
mechanism of УrowsФ in the stellar component
must be the stellar disk itself, although this hypothesis
does not rule out subsequent parallel operation of
the hydrodynamical mechanism.

 In many cases УrowsФ can be identified with
the locations of giant molecular clouds and stellar
complexes whose formation is associated with the
stellar density wave and Galactic shocks~\cite{Efremov_2010,Gusev_Efremov_2015,Efremov_2015,Bastian_Efremov_2005}.
 Chernin~\cite{Chernin_1999A&AT,Chernin_1999MNRAS}, analyzed the gas-dynamic mechanism
of the formation of the polygonal structure based
on properties 1 and 2 of the Section \ref{section:introduction}. A number
of studies~ \cite{Filistov_2012,Filistov_2015,Butenko_Khoperskov_S_2011,Khoperskov_2011_polygon_str,Khoperskov_2012_polygon_str}, used numerical simulations to
tackle the problem of the formation of straight-line
segments in the gaseous disk of the galaxy in the
gravitational potential of a smooth spiral wave, and
the results of these studies agree practically with all
the observed properties of galaxies with УrowsФ.
 %»з-за вращени€ спиральной волны звездной плотности с угловой скоростью $\Omega_s=const$, имеем радиус коротации $r_c$, на котором углова€ скорость вращени€ газа и волны звездной плотности равны $\Omega_g(r_c)=\Omega_s$.

 After the development of the global shock in numerical
models its front becomes unstable, the shock
moves out of the spiral gravitational wave and the
front becomes flat. As a result, the straight segments
of the shock forma polygonal structure with the properties
similar to those observed in real galaxies~\cite{Khoperskov_2011_polygon_str}.
Our analysis makes it possible to identify objects
where apparently only the gas-dynamicmechanism is
operating. This is primarily true for objects where the
spiral arms and УrowsФ in GALEX images extend beyond
the stellar disk. Furthermore, one to two УrowsФ can usually be found in a single arm in 2MASS images
of the old stellar disk (only in NGC~4303 a kink
can be seen in each of the three arms, whereas three or
more segments in a single arm can be recognized only
in the NGC~5156 and IC~5325 galaxies). The gasdynamic
mechanism appears to dominate in galaxies
with many УrowsФ.

 %ѕо-видимому, это обусловлено тем, что пр€мые участки наблюдаютс€ в ближнем инфракрасном H-диапазоне, в котором доминирует старое звездное население. ѕомимо NGC 4303 в работах~\cite{Melnik_Rautiainen_2013, Rautiainen_Melnik_2010} было найдено около 40 галактик с пр€мыми сегментами в изображени€х OSUBSGS: их обща€ частота встречаемости в выборке достигла 25\%, что значительно выше 7\%, полученных в~\cite{Chernin_etal_2001}. ”величение числа галактик с сегментами в~\cite{Melnik_Rautiainen_2013, Rautiainen_Melnik_2010} св€зано с использованием цифровых изображений и дополнительным применением методов обработки изображений. ¬ажно отметить, что дл€ более чем половины случаев пр€мые сегменты, наблюдаемые в B-диапазоне, можно обнаружить также в H-диапазоне. —егменты дл€ старой звездной компоненты наблюдаютс€ также в ближнем » -диапазоне на изображени€х NGC~3938 и NGC~4254 в J- и  -диапазонах~\cite{Castro-Rodriguez_2003}.

% ¬ работе~\cite{Melnik_Rautiainen_2013} было рассмотрено формирование пр€мых сегментов (<<верениц>>) в звездном галактическом диске. Ѕыли построены 2 модели, в одной из них диск в процессе динамики образует в центральной области бар, во втором случае в центре диска рождалась многорукавна€ спираль, и возникало большое количество спр€мленных участков, вращающихс€ со скоростью диска.

\section{DISCUSSION AND CONCLUSIONS}

In this study we found 276 more NGC and IC
objects with straight-line segments of spiral arms
in addition to those listed in the catalog of galaxies
with УrowsФ compiled by Chernin et al.~\cite{Chernin_etal_2001}. Together
with the 130 NGC objects from catalog~\cite{Chernin_etal_2001}, they make up a combined sample of 406 galaxies.
Unlike Chernin et al.~\cite{Chernin_etal_2001}, whose catalog is based on
an analysis of Palomar Atlas objects, we analyzed all
NGC/IC galaxies thereby ensuring the completeness
of our sample subject to additional constraints on
the observed properties. An analysis of the images
of all 7143 NGC/IC spiral galaxies allowed us to
find at least one УrowФ in the structure of the spiral
pattern of 406 objects. Thus the occurrence frequency
of galaxies with УrowsФ among the nearest objects is
of about 6\%. Of these, 77\% have central bars. A substantial
fraction (38\%) of objects with УrowsФ have
rings. Note that 13 (4\%) objects are included into the
catalog of early-type galaxies with outer rings~\cite{Kostyuk_Silchenko_2015} (nine of them are included in our sample and four
objects are listed in catalog~\cite{Chernin_etal_2001}).

Unlike objects of the catalog of Chernin et al.~\cite{Chernin_etal_2001}, where the fraction of interacting galaxies amounts to 44\% (90 objects), our sample contains only about 13.8\% (38 objects) interacting or highly asymmetric
systems. As a result, the combined sample of
all 480 objects with УrowsФ contains only 128 interacting
galaxies (27\%). Hence the hypothesis of
Chernin et al.~\cite{Chernin_etal_2001} that interacting galaxies occur almost
twice more often among galaxies with УrowsФ is
not supported by the analysis of the combined sample
of 480 objects.

Of the 276 new objects 232 galaxies (84\%) have
two-armed spiral pattern, 8\% have three arms, and
about 8\% galaxies have more than three arms. Only
12 objects (4.3\%) exhibit well-defined flocculent spiral
pattern, and the corresponding fraction is also
low in sample~\cite{Chernin_etal_2001}. УRowsФ in all arms were found
only in 59\% of our polygonal galaxies, 28\% galaxies
(78 objects) have УrowsФ only in one arm.
%“олько у 6\% галактик нашей выборки удаетс€ обнаружить вереницы по снимкам старых звезд (по данным 2MASS).
%¬ качестве примера укажем на NGC~5653, NGC~5156, NGC~6035, IC~1142, IC~1562, IC~4219, IC~4359, IC~4444, IC~4567, IC~4646, IC~4836, IC~4839, IC~5352.
%ѕодтверждаетс€ вывод о зависимости доли галактик с вереницами от относительного содержание газа (HI).

We used optical images to schematically outline
the УrowsФ in spiral arms and perform subsequent
measurements of linear and angular sizes in most of
the galaxies of our sample (93\%).  DSS images were
the main source for identifying polygonal structures
for 164 objects (59\%): DSS2 Blue (XJ+S) for 145
objects and DSS colored (composite color DSS image)
for 19 objects. SDSS DR9 color images are
used to construct images of УrowsФ for 89 objects (32\%) of our sample. In four galaxies the УrowsФ
are best recognized in optical images of the Hubble
Space Telescope (HST). Only for 19 galaxies (7\%) of our entire sample we used GALEX data as the
main images for recognizing polygonal structures,
however, УrowsФ could be recognized in ultraviolet in 51 (18\%)galaxies of our sample. In the infrared
(2MASS) straight-line segments of spiral arms can
be recognized in 18 (7\%) objects, these images usually
reveal one to two УrowsФ, whereas more such
structures can be seen in optical or ultraviolet images.
Note also that red DSS images (DSS2 Red (F+R))
reveal УrowsФ only for 13 objects of our sample.

Except for our estimate of the fraction of interacting
galaxies all other statistical properties of our
sample agree well with the results of the analysis of
the catalog of Chernin et al.~\cite{Chernin_etal_2001}.
The hypothesis about
high gas content in galaxies with УrowsФ proposed by
Chernin et al.~\cite{Chernin_etal_2001}, argues for hydrodynamical mechanism
of the formation of straight-line segments due
to the instability of powerful shocks, which agrees
with the results of numerical simulations~\cite{Khoperskov_2011_polygon_str,Butenko_Khoperskov_S_2011,Filistov_2012,Filistov_2015,Khoperskov_2012_polygon_str}. Numerical models suggest transient nature of the
formation of polygonal structures, which arise and
disappear during the evolution of the galactic disk.
The data for our sample also qualitatively suggest
that УrowsФ are non-stationary, because the spiral
pattern in about 40\% of the galaxies does not form
any regular geometric structure.

Note also that the number of УrowsФ in galaxies
of our sample is, on the average, appreciably greater
than in catalog~\cite{Chernin_etal_2001}. This, in turn, is due to the use of
images from more recent digital surveys taken in various
wavelength ranges, unlike the analysis of Chernin
et al.~\cite{Chernin_etal_2001}, which was based on a homogeneous sample
of blue images from the Palomar atlas.

% Ѕлагодарности вставл€ем через окружение {acknowledgements}

\begin{acknowledgments}
	This paper uses information contained in Hyper-
Leda astronomical database. We are grateful to the
reviewers and D. I. Makarov for valuable and useful
discussions. This work was supported by the project
within the framework of government contract from by
theMinistry of Education and Science of the Russian
Federation (project No.~2.852.2017/4.6). and partially
supported by the Russian Foundation for Basic Research
(grants no.s 15-02-06204, 15-52-12387, and
16-02-00649).M. A. Butenko thanks N. M. KuzТmin
for assistance.	
\end{acknowledgments}

\twocolumngrid
\bibliographystyle{AstroBull} % bst-файл, задающий стиль оформлени€ библиографии
\bibliography{butenko_ref_v2} % им€ bib-файла, содержащего библиографическую базу
\onecolumngrid
\clearpage

\selectlanguage{english}
\begin{center}
\large \bfseries Galaxies with Rows: New Catalog
\end{center}
\begin{center}
\bfseries M.~A.~Butenko, A.~V.~Khoperskov
\end{center}
\begin{center}
\begin{minipage}{\textwidth - 2cm}
\small
Galaxies with ``rows'' in Vorontsov-Vel'yaminov's terminology stand out among the variety of spiral galactic patterns. A characteristic feature of such objects is the sequence presence of straight segments forming a spiral arm.
 In 2001 A. Chernin and co-authors constructed a catalog of such galaxies which includes 204 objects from the Palomar Atlas.
 In this paper, we supplement the indicated catalog with 276 objects, based on an analysis of all the galaxies from the New General Catalogue and Index Catalogue.
The total number of galaxies with rows entering the NGC and IC is 406, taking into account the objects of Chernin et al. (2001).
 The use of galaxies newer images allowed us to detect more ``rows'' on average, compared with the catalog of Chernin et al.
 We did not find any significant differences between galaxies with rows and all S-galaxies from NGC/IC when comparing the most important characteristics.
We discuss two mechanisms for the formation of polygonal structures based on numerical gas-dynamic and collisionless N-body calculations that demonstrate that a spiral pattern with rows is a transient stage in the evolution of galaxies and a system with a powerful spiral structure can pass through this stage.
  The assumption of A. Chernina et al. (2001) that interacting galaxies occur twice as often among galaxies with rows is not confirmed for a combined set of 480 galaxies.
  Apparently, the presence of a central stellar bar is a favorable factor for the formation of a system of ``rows''.
\end{minipage}
\end{center}
\begin{center}
\begin{minipage}{\textwidth - 2cm}

Keywords: {galaxies: spiral---galaxies: statistics---galaxies: structure}
\end{minipage}
\end{center}
\selectlanguage{russian}

\end{document}